%% file: main.tex
\documentclass[acmsmall,screen,nonacm]{acmart}

\usepackage{placeins}  

\usepackage{amssymb}
\theoremstyle{definition}
\newtheorem{thm3}{Definition}[section]
\newtheorem{definition}[thm3]{Definition}

\AtBeginDocument{%
  \providecommand\BibTeX{{%
    \normalfont B\kern-0.5em{\scshape i\kern-0.25em b}\kern-0.8em\TeX}}}

\setcopyright{acmlicensed}
\copyrightyear{2018}
\acmYear{2018}
\acmDOI{XXXXXXX.XXXXXXX}

\acmConference[Conference acronym 'XX]{Make sure to enter the correct
  conference title from your rights confirmation emai}{June 03--05,
  2018}{Woodstock, NY}
\acmBooktitle{Woodstock '18: ACM Symposium on Neural Gaze Detection,
 June 03--05, 2018, Woodstock, NY} 
\acmISBN{978-1-4503-XXXX-X/18/06}

\input{tex/00_prelude}

\begin{document}

\title{Exo-GPU: Safe, Imperative, User-schedulable Programming for Tensor Cores}

\author{David Zhao Akeley}
\affiliation{%
  \institution{MIT CSAIL}
  \city{Cambridge}
  \state{Massachusetts}
  \country{USA}
}

\author{Yuka Ikarashi}
\affiliation{%
  \institution{MIT CSAIL}
  \city{Cambridge}
  \state{Massachusetts}
  \country{USA}
}

\author{Jonathan Ragan-Kelley}
\affiliation{%
  \institution{MIT CSAIL}
  \city{Cambridge}
  \state{Massachusetts}
  \country{USA}
}


\newif\ifshowcomments
\showcommentstrue
\ifshowcomments
\newcommand{\chengpeng}[1]{\textcolor{cyan}{[chengpeng: #1]}}
\else
\newcommand{\chengpeng}[1]{}
\fi


\begin{abstract}

\input{tex_gpu/00_abstract}
\end{abstract}

\maketitle

\input{tex_gpu/00_intro}

\input{tex_gpu/00_system_overview}

\input{tex_gpu/01_gpu_ir}
\input{tex_gpu/03_collective_analysis}

\input{tex_gpu/02_abstract_machine}
\input{tex_gpu/04_impl_results}

\input{tex_gpu/05_related_work}

\input{tex_gpu/06_limitations}
\begin{acks}
\input{tex_gpu/acknowledgements}
\end{acks}

\bibliographystyle{ACM-Reference-Format}
\bibliography{references}

\newpage
\FloatBarrier
\appendix

\input{tex_gpu/9_gpuir_types}

\input{tex_gpu/9_am_figures}



\end{document}

%% file: tex/00_prelude.tex
\makeatletter
\def\mdseries@tt{m}
\makeatother

\usepackage{amsbsy}
\usepackage{amscd}
\usepackage{amsfonts}
\usepackage{amsmath}
\usepackage{amsthm}
\usepackage{booktabs}  
\usepackage{fancybox}
\usepackage{latexsym}
\usepackage{microtype}
\usepackage{multirow}
\usepackage{stmaryrd}
\usepackage{tabularx}

\usepackage{bussproofs} 
\usepackage{cprotect}   
\usepackage{enumitem}   
\usepackage{framed}     
\usepackage{mathtools}  
\usepackage{mdframed}   
\usepackage{subcaption} 
\usepackage{caption}
\usepackage{wrapfig}
\usepackage{xcolor,colortbl}

\usepackage{pdftexcmds}
\makeatletter
\ifnum\pdf@shellescape=1
  \usepackage[finalizecache,cachedir=_minted-main]{minted}
\else
  \usepackage[frozencache,cachedir=_minted-main]{minted}
\fi
\makeatother

\usepackage{ifthen}
\usepackage{balance}
\usepackage{multicol}

\usepackage{pgfplots}
\usepackage{pgfplotstable}
\pgfplotsset{compat=1.7}

\usepackage{galois}
\usepackage{tikz-cd}
\usepackage{adjustbox}

\usepackage{tikz}
\usetikzlibrary{arrows.meta, shapes, positioning}
\usetikzlibrary{calc,intersections}
\usepackage{xcolor}
\usetikzlibrary{decorations.pathreplacing,calc}
\usetikzlibrary{tikzmark}

\usepackage{mathpartir}

\usepackage{algorithm}
\usepackage[noend]{algpseudocode}

\pgfdeclarelayer{bg}
\pgfsetlayers{bg,main}

\usepackage[normalem]{ulem}
\usepackage[inline]{aplcomments} 
\newcommenter{JRK}{0.0,0.0,1.0}
\newcommenter{glb}{1.0,0.0,0.0}
\newcommenter{Yuka}{0.8, 0.0, 0.8}
\newcommenter{Amanda}{0.0,1.0,0.0}
\newcommenter{David}{0.0,1.0,0.0}

\newcommand{\name}{Exo-GPU}
\newcommand{\ir}{GPU IR}

\newcommand{\delete}[1]{}

\definecolor{Gray}{gray}{0.85}
\newcolumntype{a}{>{\columncolor{Gray}}c}

\newcommand{\IGNORE}[1]{}

\newmintinline[code]{exolexer.py:ExoLexer -x}{fontsize=\small}

\makeatletter
\AtBeginEnvironment{minted}{\dontdofcolorbox}
\def\dontdofcolorbox{\renewcommand\fcolorbox[4][]{##4}}

\def\PYGexo@reset{\let\PYGexo@it=\relax \let\PYGexo@bf=\relax%
    \let\PYGexo@ul=\relax \let\PYGexo@tc=\relax%
    \let\PYGexo@bc=\relax \let\PYGexo@ff=\relax}
\def\PYGexo@tok#1{\csname PYGexo@tok@#1\endcsname}
\def\PYGexo@toks#1+{\ifx\relax#1\empty\else%
    \PYGexo@tok{#1}\expandafter\PYGexo@toks\fi}
\def\PYGexo@do#1{\PYGexo@bc{\PYGexo@tc{\PYGexo@ul{%
    \PYGexo@it{\PYGexo@bf{\PYGexo@ff{#1}}}}}}}
\def\PYGexo#1#2{\PYGexo@reset\PYGexo@toks#1+\relax+\PYGexo@do{#2}}

\@namedef{PYGexo@tok@w}{\def\PYGexo@tc##1{\textcolor[rgb]{0.73,0.73,0.73}{##1}}}
\@namedef{PYGexo@tok@c}{\let\PYGexo@it=\textit\def\PYGexo@tc##1{\textcolor[rgb]{0.38,0.63,0.69}{##1}}}
\@namedef{PYGexo@tok@cp}{\def\PYGexo@tc##1{\textcolor[rgb]{0.00,0.44,0.13}{##1}}}
\@namedef{PYGexo@tok@cs}{\def\PYGexo@tc##1{\textcolor[rgb]{0.38,0.63,0.69}{##1}}\def\PYGexo@bc##1{{\setlength{\fboxsep}{0pt}\colorbox[rgb]{1.00,0.94,0.94}{\strut ##1}}}}
\@namedef{PYGexo@tok@k}{\def\PYGexo@tc##1{\textcolor[rgb]{0.00,0.44,0.13}{##1}}}
\@namedef{PYGexo@tok@kp}{\def\PYGexo@tc##1{\textcolor[rgb]{0.00,0.44,0.13}{##1}}}
\@namedef{PYGexo@tok@kt}{\def\PYGexo@tc##1{\textcolor[rgb]{0.56,0.13,0.00}{##1}}}
\@namedef{PYGexo@tok@o}{\def\PYGexo@tc##1{\textcolor[rgb]{0.40,0.40,0.40}{##1}}}
\@namedef{PYGexo@tok@ow}{\def\PYGexo@tc##1{\textcolor[rgb]{0.00,0.44,0.13}{##1}}}
\@namedef{PYGexo@tok@nb}{\def\PYGexo@tc##1{\textcolor[rgb]{0.00,0.44,0.13}{##1}}}
\@namedef{PYGexo@tok@nf}{\def\PYGexo@tc##1{\textcolor[rgb]{0.02,0.16,0.49}{##1}}}
\@namedef{PYGexo@tok@nc}{\let\PYGexo@bf=\textbf\def\PYGexo@tc##1{\textcolor[rgb]{0.05,0.52,0.71}{##1}}}
\@namedef{PYGexo@tok@nn}{\let\PYGexo@bf=\textbf\def\PYGexo@tc##1{\textcolor[rgb]{0.05,0.52,0.71}{##1}}}
\@namedef{PYGexo@tok@ne}{\def\PYGexo@tc##1{\textcolor[rgb]{0.00,0.44,0.13}{##1}}}
\@namedef{PYGexo@tok@nv}{\def\PYGexo@tc##1{\textcolor[rgb]{0.73,0.38,0.84}{##1}}}
\@namedef{PYGexo@tok@no}{\def\PYGexo@tc##1{\textcolor[rgb]{0.38,0.68,0.84}{##1}}}
\@namedef{PYGexo@tok@nl}{\let\PYGexo@bf=\textbf\def\PYGexo@tc##1{\textcolor[rgb]{0.00,0.13,0.44}{##1}}}
\@namedef{PYGexo@tok@ni}{\let\PYGexo@bf=\textbf\def\PYGexo@tc##1{\textcolor[rgb]{0.84,0.33,0.22}{##1}}}
\@namedef{PYGexo@tok@na}{\def\PYGexo@tc##1{\textcolor[rgb]{0.25,0.44,0.63}{##1}}}
\@namedef{PYGexo@tok@nt}{\let\PYGexo@bf=\textbf\def\PYGexo@tc##1{\textcolor[rgb]{0.02,0.16,0.45}{##1}}}
\@namedef{PYGexo@tok@nd}{\let\PYGexo@bf=\textbf\def\PYGexo@tc##1{\textcolor[rgb]{0.33,0.33,0.33}{##1}}}
\@namedef{PYGexo@tok@s}{\def\PYGexo@tc##1{\textcolor[rgb]{0.25,0.44,0.63}{##1}}}
\@namedef{PYGexo@tok@sd}{\let\PYGexo@it=\textit\def\PYGexo@tc##1{\textcolor[rgb]{0.25,0.44,0.63}{##1}}}
\@namedef{PYGexo@tok@si}{\let\PYGexo@it=\textit\def\PYGexo@tc##1{\textcolor[rgb]{0.44,0.63,0.82}{##1}}}
\@namedef{PYGexo@tok@se}{\let\PYGexo@bf=\textbf\def\PYGexo@tc##1{\textcolor[rgb]{0.25,0.44,0.63}{##1}}}
\@namedef{PYGexo@tok@sr}{\def\PYGexo@tc##1{\textcolor[rgb]{0.14,0.33,0.53}{##1}}}
\@namedef{PYGexo@tok@ss}{\def\PYGexo@tc##1{\textcolor[rgb]{0.32,0.47,0.09}{##1}}}
\@namedef{PYGexo@tok@sx}{\def\PYGexo@tc##1{\textcolor[rgb]{0.78,0.36,0.04}{##1}}}
\@namedef{PYGexo@tok@m}{\def\PYGexo@tc##1{\textcolor[rgb]{0.25,0.63,0.44}{##1}}}
\@namedef{PYGexo@tok@gh}{\let\PYGexo@bf=\textbf\def\PYGexo@tc##1{\textcolor[rgb]{0.00,0.00,0.50}{##1}}}
\@namedef{PYGexo@tok@gu}{\let\PYGexo@bf=\textbf\def\PYGexo@tc##1{\textcolor[rgb]{0.50,0.00,0.50}{##1}}}
\@namedef{PYGexo@tok@gd}{\def\PYGexo@tc##1{\textcolor[rgb]{0.63,0.00,0.00}{##1}}}
\@namedef{PYGexo@tok@gi}{\def\PYGexo@tc##1{\textcolor[rgb]{0.00,0.63,0.00}{##1}}}
\@namedef{PYGexo@tok@gr}{\def\PYGexo@tc##1{\textcolor[rgb]{1.00,0.00,0.00}{##1}}}
\@namedef{PYGexo@tok@ge}{\let\PYGexo@it=\textit}
\@namedef{PYGexo@tok@gs}{\let\PYGexo@bf=\textbf}
\@namedef{PYGexo@tok@gp}{\let\PYGexo@bf=\textbf\def\PYGexo@tc##1{\textcolor[rgb]{0.78,0.36,0.04}{##1}}}
\@namedef{PYGexo@tok@go}{\def\PYGexo@tc##1{\textcolor[rgb]{0.53,0.53,0.53}{##1}}}
\@namedef{PYGexo@tok@gt}{\def\PYGexo@tc##1{\textcolor[rgb]{0.00,0.27,0.87}{##1}}}
\@namedef{PYGexo@tok@err}{\def\PYGexo@bc##1{{\setlength{\fboxsep}{\string -\fboxrule}\fcolorbox[rgb]{1.00,0.00,0.00}{1,1,1}{\strut ##1}}}}
\@namedef{PYGexo@tok@kc}{\let\PYGexo@bf=\textbf\def\PYGexo@tc##1{\textcolor[rgb]{0.00,0.44,0.13}{##1}}}
\@namedef{PYGexo@tok@kd}{\let\PYGexo@bf=\textbf\def\PYGexo@tc##1{\textcolor[rgb]{0.00,0.44,0.13}{##1}}}
\@namedef{PYGexo@tok@kn}{\let\PYGexo@bf=\textbf\def\PYGexo@tc##1{\textcolor[rgb]{0.00,0.44,0.13}{##1}}}
\@namedef{PYGexo@tok@kr}{\let\PYGexo@bf=\textbf\def\PYGexo@tc##1{\textcolor[rgb]{0.00,0.44,0.13}{##1}}}
\@namedef{PYGexo@tok@bp}{\def\PYGexo@tc##1{\textcolor[rgb]{0.00,0.44,0.13}{##1}}}
\@namedef{PYGexo@tok@fm}{\def\PYGexo@tc##1{\textcolor[rgb]{0.02,0.16,0.49}{##1}}}
\@namedef{PYGexo@tok@vc}{\def\PYGexo@tc##1{\textcolor[rgb]{0.73,0.38,0.84}{##1}}}
\@namedef{PYGexo@tok@vg}{\def\PYGexo@tc##1{\textcolor[rgb]{0.73,0.38,0.84}{##1}}}
\@namedef{PYGexo@tok@vi}{\def\PYGexo@tc##1{\textcolor[rgb]{0.73,0.38,0.84}{##1}}}
\@namedef{PYGexo@tok@vm}{\def\PYGexo@tc##1{\textcolor[rgb]{0.73,0.38,0.84}{##1}}}
\@namedef{PYGexo@tok@sa}{\def\PYGexo@tc##1{\textcolor[rgb]{0.25,0.44,0.63}{##1}}}
\@namedef{PYGexo@tok@sb}{\def\PYGexo@tc##1{\textcolor[rgb]{0.25,0.44,0.63}{##1}}}
\@namedef{PYGexo@tok@sc}{\def\PYGexo@tc##1{\textcolor[rgb]{0.25,0.44,0.63}{##1}}}
\@namedef{PYGexo@tok@dl}{\def\PYGexo@tc##1{\textcolor[rgb]{0.25,0.44,0.63}{##1}}}
\@namedef{PYGexo@tok@s2}{\def\PYGexo@tc##1{\textcolor[rgb]{0.25,0.44,0.63}{##1}}}
\@namedef{PYGexo@tok@sh}{\def\PYGexo@tc##1{\textcolor[rgb]{0.25,0.44,0.63}{##1}}}
\@namedef{PYGexo@tok@s1}{\def\PYGexo@tc##1{\textcolor[rgb]{0.25,0.44,0.63}{##1}}}
\@namedef{PYGexo@tok@mb}{\def\PYGexo@tc##1{\textcolor[rgb]{0.25,0.63,0.44}{##1}}}
\@namedef{PYGexo@tok@mf}{\def\PYGexo@tc##1{\textcolor[rgb]{0.25,0.63,0.44}{##1}}}
\@namedef{PYGexo@tok@mh}{\def\PYGexo@tc##1{\textcolor[rgb]{0.25,0.63,0.44}{##1}}}
\@namedef{PYGexo@tok@mi}{\def\PYGexo@tc##1{\textcolor[rgb]{0.25,0.63,0.44}{##1}}}
\@namedef{PYGexo@tok@il}{\def\PYGexo@tc##1{\textcolor[rgb]{0.25,0.63,0.44}{##1}}}
\@namedef{PYGexo@tok@mo}{\def\PYGexo@tc##1{\textcolor[rgb]{0.25,0.63,0.44}{##1}}}
\@namedef{PYGexo@tok@ch}{\let\PYGexo@it=\textit\def\PYGexo@tc##1{\textcolor[rgb]{0.38,0.63,0.69}{##1}}}
\@namedef{PYGexo@tok@cm}{\let\PYGexo@it=\textit\def\PYGexo@tc##1{\textcolor[rgb]{0.38,0.63,0.69}{##1}}}
\@namedef{PYGexo@tok@cpf}{\let\PYGexo@it=\textit\def\PYGexo@tc##1{\textcolor[rgb]{0.38,0.63,0.69}{##1}}}
\@namedef{PYGexo@tok@c1}{\let\PYGexo@it=\textit\def\PYGexo@tc##1{\textcolor[rgb]{0.38,0.63,0.69}{##1}}}

\makeatother

\numberwithin{equation}{section}
\theoremstyle{plain} 

\theoremstyle{definition} 
 
\theoremstyle{remark}

\newcommand{\Expr}{\mathsf{Expr}}
\newcommand{\Stmt}{\mathsf{Stmt}}
\newcommand{\Proc}{\mathsf{Proc}}

\newcommand{\qc}[1]{\left\llbracket#1\right\rrbracket}

\ifdefined\Wr
\renewcommand{\Wr}{\mathbf{Wr}} 
\else
\newcommand{\Wr}{\mathbf{Wr}} 
\fi

\let\YesAppendix\relax 

\ifdefined\YesAppendix
\newcommand{\appendixref}[1]{\ref{#1}}
\else 
\newcommand{\appendixref}[1]{%
\ifthenelse{\equal{#1}{appendix:core-lang}}{A}{}%
\ifthenelse{\equal{#1}{appendix:ternary-logic}}{B}{}%
\ifthenelse{\equal{#1}{appendix:global-dataflow}}{C}{}%
\ifthenelse{\equal{#1}{appendix:location-set-membership}}{D}{}%
\ifthenelse{\equal{#1}{appendix:effect-extraction}}{E}{}%
\ifthenelse{\equal{#1}{appendix:context-analysis}}{F}{}%
\ifthenelse{\equal{#1}{appendix:gemmini_library}}{G}{}%
}
\fi

%% file: tex_gpu/00_abstract.tex

Modern GPUs require not only SIMT-style parallelism but also software-managed concurrency between compute and data movement to reach maximum performance. Performance engineers must reason about subdividing work into the hierarchy of computation resources (threads, warps, warpgroups, blocks, clusters), and, in many cases, also must use asynchronous tensor core and memcpy instructions on different levels of the memory hierarchy (registers, tensor core accumulators, shared memory, global memory).
Unlike CPUs, where out-of-order execution is managed by hardware and hidden from programmers, GPUs expose explicit instruction reordering to software through these asynchronous instructions.
Well-established GPU programming languages generally offer either direct low-level control without safety guarantees (e.g., CUDA C++ inline assembly or intrinsics) or easier-to-analyze, high-level abstractions (e.g., Triton’s tile-based model) that hide asynchronous instructions in the compiler backend, which may prevent performance engineers from maximizing performance by tuning critical details.

We propose Exo-GPU, an imperative, low-level language that creates minimal abstraction over CUDA.
Our key idea is to treat parallelism and synchronization as mere annotations on sequential code rather than as fundamental control flow primitives, enabling verification that these constructs do not alter the program semantics.
The benefit is two-fold: programmers can reason about code without hidden control flow or mutation, while allowing the \name{} compiler to verify \emph{sequential-parallel equivalence}---guaranteeing that parallel execution is functionally equivalent to its sequential interpretation.
We used Exo-GPU to author GEMM kernels for the H100 GPU, using wgmma, TMA, and split-k.
Our kernels achieved over 80\% of theoretical peak on large problem sizes, in some cases outperforming the vendor-provided CUBLAS library.

%% file: tex_gpu/00_intro.tex
\vspace{-0.6em}
\section{Introduction}
\vspace{-0.3em}

Starting with Tensor Cores introduced in NVIDIA's Volta generation of GPUs \cite{V100}, the menu of tensor-specialized features in NVIDIA GPUs has expanded to encompass automated bulk movement of multidimensional data, distributed shared memory, and new asynchronous tensor cores supporting large-scale (up to $64 \times 256$) matrix accumulation in a single instruction \cite{ptx}.
This poses ever-increasing challenges for authoring correct, top-performing CUDA code.
Not only must algorithms be rewritten to migrate work from traditional SIMT-style code to target these specialized instructions, but also, these instructions expose explicit out-of-order instruction execution to the programmer, moving synchronization that was traditionally the responsibility of the hardware (e.g., through scoreboarding) to the programmer.

High-level kernel programming languages, such as Triton \cite{triton}, provide a popular approach for taming this complexity.
Triton's tile-based, automatically-parallelized programming model reduces the risk of programmer error by moving responsibility for correct usage of bulk data movement and synchronization instructions into the compiler.
However, in practice, groundbreaking algorithms, such as the original FlashAttention 3 \cite{shah2024flashattention3fastaccurateattention}, are often written directly in CUDA C++ for greater control.
On the one hand, CUDA C++ provides direct control over features such as asynchronous data movement and warp specialization required for optimal GPU usage.
On the other hand, CUDA C++ exposes programmers to a variety of potential bugs, which we broadly categorize as:
\vspace{-0.2em}
\begin{enumerate}
    \item \emph{Dataflow Logic Bugs}: Program optimization usually requires changing the storage or computation order of intermediate values, e.g. caching data in a faster memory, or reordering instructions to hide latency.
    A flawed optimization, such as an indexing bug in copying a tile, or exchanging two statements that don't commute safely, can change the functional behavior of the program.
    \item \emph{Instruction Logic Bugs}: Accelerator instructions won't work as intended if not used as specified by the hardware vendor.
    For example, \texttt{wgmma} tensor core instructions require data in a certain input format and  uniform execution by a full warpgroup (128 aligned threads).
    \item \emph{Inter-Thread Synchronization Bugs}: Garden-variety synchronization bugs involve two threads accessing the same memory location (with one access being a write) without ordering enforced by synchronization.
    \item \emph{Intra-Thread Synchronization Bugs}: If one thread issues an asynchronous (async) instruction, and that \emph{same thread} later issues another instruction (async or not) that operates on the same memory location, overlapped or reordered instruction execution can undermine correctness.
\end{enumerate}
\vspace{-0.3em}

Our goal is to provide a programming model that gives CUDA-like control \emph{without} accepting the risk of these bugs as a fact of life.
Existing CUDA C++ libraries such as CuTe \cite{cute} and ThunderKittens \cite{thunderkittens} provide close-to-the-metal abstractions that minimize the risk of logic bugs, particularly data movement and instruction input format bugs; however, none of them support large-scale program analysis and synchronization checking of arbitrary input programs.

As a starting point for the \name{} language design, we postulate that programs with \emph{sequential semantics} are far easier to analyze, both by humans and by compilers.
We define program behavior using standard sequential semantics, treating language features for parallelism and synchronization--including parallel for-loops across CUDA thread hierarchies and barrier statements--simply as \emph{annotations} on sequential code.
This design provides two key benefits: programmers can intuitively reason about their code without hidden control flow or side effects, and more importantly, compilers can verify whether synchronizations are legal; that is, they can verify these annotations do not alter the program behavior relative to its baseline sequential interpretation.

The \name{} program, interpreted sequentially, serves as as the semantic baseline---defining \emph{what} the user intends to compute. Parallelism annotations specify \emph{which threads} execute which statement instances, while synchronization constructs specify \emph{how} the user intends to uphold baseline behavior.
This raises a question: what degree of safety and control does \name{} provide?
We define safety as \emph{sequential-parallel equivalence}--ensuring that the parallel interpretation of a program matches the sequential one.
Parallelism is explicitly controlled by the programmer's annotations, rather than being hidden and automated by the compiler.
The \name{} compiler's role is limited to \emph{verifying} that the user-provided synchronization ensures equivalence.




Through collective analysis (Section~\ref{sec:collective-analysis}), we annotate each statement with a \emph{collective tiling}, specifying which thread(s) execute each statement instance.
This enables static analysis of the thread assignment counts, ensuring instructions with thread convergence requirements (e.g. warp MMA) are used correctly.
To ensure sequential-parallel equivalence, we simulate execution on an \emph{abstract machine}  (Section~\ref{sec:abstract-machine}).
The \name{} program converts to an Abstract Machine program where data variables store access histories (reads and mutations) with their thread and async instruction \emph{visibility}, rather than numeric values.
Synchronization statements become meaningful only in the abstract machine, conditionally expanding the visibility of all prior recorded accesses.




We built \name{} as an extension to Exo \cite{exo, exo2}, a user-schedulable language providing composable, checked rewrite rules for transforming sequential programs.
Exo lacks GPU support, and we significantly extended its frontend language design, static analysis, semantics, and codegen to enable safe parallel program transformation.
Treating parallel constructs as annotations over a sequential semantic baseline enabled us to reuse Exo's over 60 verified rewrites as-is, which allowed us to transform a simple sequential GEMM program into a complex GEMM kernel for the NVIDIA Hopper architecture, with every step of transformation verified via a \emph{chain of equivalence}: Exo verified equivalence between the original $p_1$ and the final $p_n$ while ignoring all the synchronization constructs, and \name{} verified functional equivalence between the parallel and sequential interpretations of $p_n$, checking the legality of the parallel and synchronous annotations (Section~\ref{sec:optimization}).
Combining Exo's pre-existing checked rewrite rules with \name{}'s new language features and synchronization checking, we created a GEMM kernel achieving over 80\% of theoretical peak for large problem sizes, in some cases outperforming the vendor-supplied CUBLAS library (Section~\ref{sec:results}).

%% file: tex_gpu/00_system_overview.tex
\section{Examples and System Overview}
\label{sec:examples-system-overview}

\subsection{\name{} Language Overview}
\label{sec:language-overview}

\name{} is built on fundamental design goals of safe, performance transparent, and imperative GPU programming.
Unlike other GPU abstractions that hide complexity, we make threads and most CUDA instructions explicit in the IR. 
We introduce three first-class frontend language features derived from our design principles: (i) parallel loops that explicitly map work to blocks, warps, or threads, (ii) a distributed memory model with sharding and explicit memory space annotations, and (iii) explicit synchronization, including fences, split-barriers, and asynchronous operations.

\textbf{Parallel loops.}
\name{} exposes three kinds of loops.\\
(1) \texttt{seq}$(c_{lo},c_{hi})$ is a sequential loop that iterates $i$ from $c_{lo}$ to $c_{hi}-1$.\\
(2) \texttt{cuda\_tasks}$(lo,hi)$ distributes iterations across CUDA clusters; on pre-\texttt{sm\_90} (H100) GPUs a cluster coincides with a single CTA (thread block).\\
(3) \texttt{cuda\_threads}$(lo,hi,\; \texttt{unit}=\tau_u)$ assigns iterations to subsets of threads inside the current cluster.\\
The \texttt{unit} takes a parameter $\tau_u$ that specifies the shape of the participating thread set but not concrete indices—for example, $\tau_u=\,$\texttt{cuda\_warp} denotes contiguous 32‑thread groups with thread IDs \(0\text{--}31\), \(32\text{--}63\), \(64\text{--}95\), etc.\ (inclusive), while $\tau_u=\,$\texttt{cuda\_thread} denotes a single thread (see Figure~\ref{fig:CollUnit}).
%



\begin{wrapfigure}{r}{0.34\textwidth}
\vspace{-5mm}
\begin{minted}[fontsize=\footnotesize,breaklines]{python}
for blk in cuda_tasks(0, 37):
  # scope-1 (CTA-scope)
  for w in cuda_threads(0, 4,
               unit=cuda_warp):
    # scope-2 (warp-scope)
    for t in cuda_threads(0, 32,
               unit=cuda_thread):
      # scope-3 (thread-scope)
\end{minted}
\vspace{-5mm}
\end{wrapfigure}
%
%
%
%

Each loop body, conditional branch, \texttt{with}-block, or procedure defines a \emph{program scope}.
As part of \emph{collective analysis} (Section~\ref{sec:collective-analysis}), we annotate each scope with a \emph{collective tiling} $\omega \in \Omega$.
This describes a mapping between the control variable values and the \emph{thread collective} (set of threads) assigned to execute statement instances in the scope (i.e. which threads execute which loop iterations).
We say that a scope is a $\tau_u$-scope when $\tau_u$ accurately describes the thread sets produced by the mapping.
In the inset figure, the outermost parallel loop \texttt{cuda\_tasks(0,37)} creates \texttt{scope‑1} (which is a cluster scope, and, assuming \texttt{clusterDim=1}, also a CTA scope); nesting \texttt{cuda\_threads(..., unit=\texttt{cuda\_warp})} creates \texttt{scope‑2} (warp scope); and nesting \texttt{cuda\_threads(..., unit=\texttt{cuda\_thread})} creates \texttt{scope‑3} (single‑thread scope). 
Statement instances in \texttt{scope‑3} are executed by one thread.

\textbf{Distributed Memory:}
\name{} exposes explicit allocation statements for data and barrier variables, annotated by memory type, e.g. \texttt{x : f32[8] @ CudaRmem} allocates a data array \texttt{x} of type \texttt{f32} and size 8 in GPU register memory.
Allocations can be \emph{distributed}: a single logical object may be sharded across threads at multiple CUDA levels (e.g., per-thread register shards or per-CTA-in-cluster shared-memory shards). Threads must access only their owned shards except via explicit communication instructions (e.g., warp shuffles) or TMA multicast.
This sharding is not annotated explicitly at the allocation, but deduced from the usage pattern of the variable; all deductions must be consistent, as enforced by distributed memory analysis (Section~\ref{sec:distributed-memory-analysis}).

\begin{wrapfigure}{r}{0.34\textwidth}
\vspace{-5mm}
\begin{minted}[fontsize=\footnotesize,breaklines]{python}
tmp: f32 @ CudaRmem  # Register
for i in cuda_threads(0, 32,
           unit=cuda_thread):
  tmp = A[i]
for i in cuda_threads(0, 32,
           unit=cuda_thread):
  B[i] = tmp
\end{minted}
\vspace{-5mm}
\end{wrapfigure}
%
\newpage  
Inset right illustrates the motivation for this design.
If the first loop was executed sequentially, \texttt{tmp} would hold \texttt{A[31]} after the first loop, so the second loop broadcasts that value to every \texttt{B[i]}.
Languages like CUDA that implicitly duplicate local variables per thread would interpret the same code as a vector copy (\texttt{B[i] = A[i]}).
While not necessarily error-prone, this design would be inappropriate for our goal of defining program behavior with sequential semantics.
Hence, in \name{}, storing a temporary value per-thread requires a sharded allocation of \texttt{tmp}; as written, the program is rejected by the analysis because a register-resident \texttt{tmp} cannot be read by different threads without (for example) a warp shuffle.
If a user intends to express a vector copy (\texttt{B[i] = A[i]}) through register \texttt{tmp}, \texttt{tmp} requires explicit \emph{distributed} allocation per thread (\texttt{tmp:f32[32]}) and the accesses must be per-thread (\texttt{tmp[i]}).


\textbf{Explicit Synchronization:}
Many GPU instructions require additional explicit synchronization, but overly conservative barriers can cancel the performance benefits of concurrency. \name{} provides two forms of synchronization: a blocking barrier via \texttt{Fence}, and split barriers via paired \texttt{Arrive}/\texttt{Await}. As with other statements, all synchronization statements execute in program order.

\begin{wrapfigure}{r}{0.5\textwidth}
\vspace{-5mm}
\begin{minted}[fontsize=\footnotesize,breaklines]{python}
# Code at CTA-scope, with blockDim=128
buf: f32[128] @ CudaSmemLinear
for tid in cuda_threads(0, 128, unit=cuda_thread):
  buf[tid] = gmem[tid]
# __syncthreads
Fence(cuda_in_order,cuda_in_order)
for tid in cuda_threads(0, 128, unit=cuda_thread):
  accum: f32 # Each thread sums up buf[...]
  accum = 0
  for i in seq(0, 128):
    accum += buf[i]
\end{minted}
\vspace{-5mm}
\end{wrapfigure}
%
%
%
A \texttt{Fence} causes all threads in the current thread collective (warp, CTA, or cluster) to rendezvous. At warp scope it maps to \texttt{\_\_syncwarp}; at CTA scope it maps to \texttt{\_\_syncthreads} (as shown in the right-hand code inset).
A \texttt{Fence} takes two synchronization–timeline parameters, $\tau^{\text{pre}}_s$ and $\tau^{\text{post}}_s$ (Figure~\ref{fig:SyncTL-def}).
These timelines specify which memory operations (sync and/or async) before and after the fence must be ordered, and the compiler emits the minimal combination of synchronization and memory-fence instructions required to satisfy them.
The \texttt{cuda\_in\_order} timeline specifies that the effects of async instructions are \emph{not} synchronized; only typical synchronous instruction effects are.
If we were to, for example, replace the loads into shared memory with \texttt{cp.async} operations,
the first parameter must be set to \texttt{Sm80\_cp\_async} so that the copy operations preceding the fence are completed before execution proceeds.


\begin{wrapfigure}{r}{0.33\textwidth}
\vspace{-4mm}
\begin{minted}[fontsize=\footnotesize,breaklines]{python}
for tid in cuda_threads(0, 128,
             unit=cuda_thread):
  ...
bar: barrier @ CudaMbarrier
Arrive(cuda_in_order) >> bar
# ... insert work here
Await(bar, cuda_in_order)
for tid in cuda_threads(0, 128,
             unit=cuda_thread):
  ...
\end{minted}
\vspace{-5mm}
\end{wrapfigure}
%
%
Split barriers use a shared \texttt{barrier}-typed variable $z$, declared as \texttt{$z$: barrier @ $\pi_b$}, where $\pi_b$ selects the completion mechanism  (commit group, mbarrier, or cluster sync).
Programmers express a split barrier statement pair with \texttt{Arrive($\tau^{\text{pre}}_s$) >}\texttt{> $z$} and \texttt{Await($z$, $\tau^{\text{post}}_s$)}.
The statement pairing is determined by the shared barrier variable $z$. 
A matching pair orders memory operations prior to \texttt{Arrive} (filtered by $\tau^{\text{pre}}_s$) with those after \texttt{Await} (filtered by $\tau^{\text{post}}_s$).
The right-hand inset illustrates replacing the earlier \texttt{Fence} example with an \texttt{mbarrier}-based split barrier.









\subsection{\name{} Examples with Erroneous Synchronization}
\label{sec:sync-errors}



To motivate our language design, we 
build up a series of increasingly subtle, hard-to-identify inter- and intra-thread synchronization bugs; 
all would be rejected by synchronization checking (Section~\ref{sec:abstract-machine}).
Variables prefixed with \texttt{some\_*} are allocated outside the code snippet, \texttt{some\_in\_order\_op} represents non-async work, and \texttt{example\_tma\_multicast} and \texttt{example\_wgmma} are simplified placeholders for  actual TMA (tile copy) and wgmma (tensor core) instructions.


The first example illustrates an inter-thread synchronization bug involving a cluster with 2 CTAs:

\begin{minted}[fontsize=\footnotesize,breaklines]{python}
B: f32[2, 256, 32] @ CudaSmemLinear # Distributed mem: B[0,:,:] on CTA 0; B[1,:,:] on CTA 1
for cta in cuda_threads(0, 2, unit=cuda_cta_in_cluster):
    some_in_order_op(B[cta, :, :])
    Fence(cuda_in_order, cuda_in_order) # __syncthreads
example_tma_multicast(B[:,:,:], some_gmem[:,:])  # Copy some_gmem[:,:] to B[0,:,:] and B[1,:,:]
\end{minted}
\noindent
The \texttt{example\_tma\_multicast} instruction is implemented by threads in the 2 CTAs cooperating to fill each others' shared memory; however, this cooperation does not entail implicit synchronization between the two CTAs.
Since the \texttt{Fence} statement, as placed in CTA-scope, only synchronizes threads within a CTA, it is possible that threads in CTA 0 proceed to the multicast instruction, filling the SMEM of CTA 1 (\texttt{B[1,:,:]}), while threads in CTA 1 are still reading from it while executing \texttt{some\_in\_order\_op}.
Although the TMA instruction is async, this is not relevant to the bug illustrated here, as the TMA appears after the in-order operations upon which it races with.

The next example illustrates an intra-thread synchronization bug:

\begin{minted}[fontsize=\footnotesize,breaklines]{python}
D: f32[2, 2, 64, 256] @ Sm90_RmemMatrixD(64, 256)  # Distributed: D[x,y,:,:] on CTA x, warpgroup y
for cta in cuda_threads(0, 2, unit=cuda_cta_in_cluster):
    for wg in cuda_threads(0, 2, unit=cuda_warpgroup):
        example_wgmma(D[cta, wg, :, :], some_A[:, :], some_B[:, :])  # D += matmul(A, B)
        some_in_order_op(D[cta, wg, :, :])
\end{minted}
\noindent
This code parallelizes across CTAs and warpgroups, with each warpgroup accessing only its own shard \texttt{D[cta, wg, :, :]}.
Nevertheless, the code is flawed, as the async \texttt{example\_wgmma} instruction isn't waited-for before \texttt{some\_in\_order\_op} attempts to access its output.

A more subtle bug arises when async instructions interact with inter-thread synchronization:

\begin{tikzpicture}[remember picture]
    \node[anchor=north west,inner sep=0] (code) at (0,0) {%
\begin{minipage}[t]{\textwidth}
\begin{minted}[fontsize=\footnotesize,breaklines,escapeinside=§§]{python}
B: f32[2, 256, 32] @ Sm90_SmemSwizzled(128)        # Distributed: B[x,:,:] on CTA x
D: f32[2, 2, 64, 256] @ Sm90_RmemMatrixD(64, 256)  # Distributed: D[x,y,:,:] on CTA x, warpgroup y
cg: barrier[2, 2] @ CudaCommitGroup                # Distributed: cg[x,y] for CTA x, warpgroup y
for cta in cuda_threads(0, 2, unit=cuda_cta_in_cluster):
    for wg in cuda_threads(0, 2, unit=cuda_warpgroup):
        §\raisebox{0.25\baselineskip}{\tikzmark{bug3_B_io}}§some_in_order_op_1(B[cta, :, :])
        example_wgmma(D[cta, wg, :, :], some_A[:,:], §\textbf{B[cta\tikzmark{bug3_wgmma}, :, :]}§)  # D += matmul(A, B)
        Arrive(wgmma_async) >> cg[cta, wg] §\raisebox{0.25\baselineskip}{\tikzmark{bug3_cg_Arrive}}§
cs: barrier @ CudaClusterSync
§\raisebox{0.25\baselineskip}{\tikzmark{bug3_cs_Arrive}}§Arrive(cuda_in_order) >> cs §\raisebox{0.25\baselineskip}{\tikzmark{bug3_cs_Arrive_right}}§
for cta in cuda_threads(0, 2, unit=cuda_cta_in_cluster):
    for wg in cuda_threads(0, 2, unit=cuda_warpgroup):
        Await(cg[cta, wg], cuda_in_order, 0) §\raisebox{0.25\baselineskip}{\tikzmark{bug3_cg_Await}}§
        some_in_order_op_2(D[cta, wg, :, :]) §\raisebox{0.25\baselineskip}{\tikzmark{bug3_D_io}}§
§\raisebox{0.25\baselineskip}{\tikzmark{bug3_cs_Await}}§Await(cs, cuda_in_order, 0)
§\raisebox{0.25\baselineskip}{\tikzmark{bug3_tma}}§example_tma_multicast(B[:,:,:], some_gmem[:,:])  # Copy some_gmem[:,:] to B[0,:,:] and B[1,:,:]
\end{minted}
\end{minipage}};
\newcommand{\bugYFix}{-2.4mm}
\coordinate (bugFixBIo) at ([yshift=\bugYFix]pic cs:bug3_B_io);
\coordinate (bugFixWgmma) at ([yshift=\bugYFix]pic cs:bug3_wgmma);
\coordinate (bugFixCgArrive) at ([yshift=\bugYFix]pic cs:bug3_cg_Arrive);
\coordinate (bugFixCsArrive) at ([yshift=\bugYFix]pic cs:bug3_cs_Arrive);
\coordinate (bugFixCsArriveRight) at ([yshift=\bugYFix]pic cs:bug3_cs_Arrive_right);
\coordinate (bugFixCgAwait) at ([yshift=\bugYFix]pic cs:bug3_cg_Await);
\coordinate (bugFixDIo) at ([yshift=\bugYFix]pic cs:bug3_D_io);
\coordinate (bugFixCsAwait) at ([yshift=\bugYFix]pic cs:bug3_cs_Await);
\coordinate (bugFixTma) at ([yshift=\bugYFix]pic cs:bug3_tma);
\draw[-,thick,blue,overlay] (bugFixBIo) to[out=180, in=180] (bugFixCsArrive);
\draw[-,thick,blue,overlay] (bugFixCsArrive) to[out=180, in=180] (bugFixCsAwait);
\draw[-,thick,blue,overlay] (bugFixCsAwait) to[out=180, in=180] (bugFixTma);
\draw[-,thick,blue,overlay] (bugFixWgmma) to[out=270, in=0] (bugFixCgArrive);
\draw[-,thick,blue,overlay] (bugFixCgArrive) to[out=0, in=0] (bugFixCgAwait);
\draw[-,thick,blue,overlay] (bugFixCgAwait) to[out=0, in=0] (bugFixDIo);
\draw[->,thick,dotted,red,overlay] (bugFixWgmma) to[out=270, in=0] (bugFixCsArriveRight);
\draw[->,thick,dotted,red,overlay] (bugFixCgAwait) to[out=90, in=0] (bugFixCsArriveRight);
\end{tikzpicture}
The statements connected in blue on the left illustrate safe synchronization in a situation similar to our first example: the \texttt{Arrive/Await} pair using \texttt{cs} (cluster sync) ensures all accesses to \texttt{B} by \texttt{some\_in\_order\_op\_1} finish before any thread in the cluster overwrites \texttt{B} using TMA.
The statements connected in blue on the right illustrate safe synchronization in a situation similar to our second example: the \texttt{Arrive/Await} pair using \texttt{cg} (commit group) ensures each warpgroup waits for its \texttt{example\_wgmma} to finish before accessing \texttt{D[cta, wg...]} in \texttt{some\_in\_order\_op\_2}.

However, the bolded access to \textbf{\texttt{B}} by \texttt{example\_wgmma} is unsafe.
Dotted red arrows illustrate synchronization that does \emph{not} happen.
Because the \texttt{Arrive} on \texttt{cs} uses the timeline \texttt{cuda\_in\_order}, the completion of the \texttt{Arrive} does not depend on the prior async wgmma instructions to retire.
Meanwhile, the \texttt{Await} on \texttt{cg} (commit group) occurs too late for it to interact transitively with the prior \texttt{cs} (cluster sync) \texttt{Arrive}.
Therefore, while all \emph{non-async} accesses to \texttt{B} will retire before any thread proceeds to the TMA multicast, this isn't the case for the async wgmma instructions, whose \texttt{B} operand could be overwritten unexpectedly by another CTA multicasting to its input memory.
Subtle bugs like this can be difficult to spot in optimized CUDA code. 

\subsection{Optimization Process and Equivalence Checking}
\label{sec:optimization}

\name{} programmers can either directly write optimized object code or apply a sequence of scheduling transformations to simple initial code.
Both approaches grant full control over performance-critical details.
However, the scheduling approach offers significant practical advantages for performance engineering: 
optimization strategies become reusable functions across similar kernels, complex transformations decompose into sequences of verified primitives that enable rapid iteration with correctness guarantees, and the compiler automatically handles complex index calculations for tiling strategies---especially valuable for GPU optimization.
%
%
Below, we describe \name{}'s scheduling-based optimization process.

%

The scheduling flow in \name{} begins with a simple initial procedure $p_1$.
The programmer issues a series of \emph{rewrite operations} $R_1$ ... $R_{N-1}$ to create transformed procedures $p_2 ... p_N$.

\noindent
\begin{tabular}{l c c c c c c c c}
Rewrites: & $p_1$ & $\xrightarrow{R_1}$ & $p_2$ & ... & $\xrightarrow{R_{N-1}}$ & $p_N$ & \footnotesize $\text{(\name{} checks)}$ & $p_N$ \\[2pt]
Behavior: & $\textsf{seq}\qc{p_1}$ & $\equiv$ & $\textsf{seq}\qc{p_2}$ & ... & $\equiv$ & $\textsf{seq}\qc{p_N}$ & $\equiv$ & $\textsf{par}\qc{p_N}$
\end{tabular}

\vspace{2mm}
\noindent
Exo's scheduling rewrite verifies functional equivalence of $\textsf{seq}\qc{p_1} ... \textsf{seq}\qc{p_N}$ under sequential semantics (denoted $\mathsf{seq}\qc{...}$).
\name{} adds new scheduling operations for parallel and synchronization constructs, but these non-semantics-altering annotations are ignored under $\mathsf{seq}\qc{...}$ checks.
Then \name{} performs additional verification on the final program $p_N$.
Specifically, synchronization checking ensures $\textsf{seq}\qc{p_N} \equiv \textsf{par}\qc{p_N}$, where $\mathsf{par}\qc{...}$ denotes parallel semantics.
Transitively, this completes the \emph{chain of equivalence} from $\textsf{seq}\qc{p_1} \equiv ... \equiv \textsf{seq}\qc{p_N} \equiv \textsf{par}\qc{p_N}$, guaranteeing that $p_1$'s behavior is preserved throughout all transformations. 

Equivalence checking relies on three categories of reasoning:






\textbf{Exo Rewrites:}
Exo's existing rewrites implement dependency analysis-based equivalence checking, which guards against \emph{dataflow logic bugs}.
For example, \texttt{reorder\_stmts}, which reorders two statements $s_1;s_2$, is only permitted when the effects of the statements commute.


\textbf{Instruction Substitution:}
Explicit instruction selection rewrite (\texttt{replace}) lets programmers substitute a code block with an accelerator instruction call.
\name{} models instructions via their \emph{behavior} (sequential semantics), per-parameter \emph{memory type} (physical memory and layout), expected collective unit $\tau_u$, and async annotations.
Exo's unification verifies that the replaced code block's behavior matches the instruction, while \name{}'s static analysis ensures appropriate threads execute it (Section~\ref{sec:distributed-memory-analysis}); together preventing \emph{instruction logic errors}.


\textbf{\name{} Synchronization Check:}
Since parallel loops and synchronization statements (\texttt{Fence}, \texttt{Arrive}, and \texttt{Await}) are ignored during sequential checks, synchronization checking must ensure that their usage is valid and does not alter program behavior or introduce race conditions.
We verify this via interpretation on the Abstract Machine (Section~\ref{sec:abstract-machine}), which protects the program against both \emph{inter- and intra-thread synchronization bugs} for selected concrete problem sizes.





%% file: tex_gpu/01_gpu_ir.tex
\vspace{-0.5em}
\section{GPU IR}
\label{sec:gpu-ir}



\ir{} is the frontend language of \name{}, and examples in Section~\ref{sec:examples-system-overview} are written in \ir{}.

\vspace{-0.5em}
\subsection{Variables and Expressions}
\label{sec:gpuir-exprs}
\input{tex_gpu/grammar_gpu_ir_exprs}

Figure~\ref{fig:gpuir-expression-syntax-def} defines the expression syntax of \ir{}.
\ir{} explicitly distinguishes between control, data, barrier, and warp expressions and is a static control program. We denote the sets of data, control, barrier, and warp variables by $\mathbb{X}$, $\mathbb{Y}$, $\mathbb{B}$, $\mathbb{W}$, respectively. $x \in \mathbb{X}$ is a data variable, $y \in \mathbb{Y}$ is a control variable, $z \in \mathbb{B}$ is a barrier variable, and $r \in \mathbb{W}$ is a warp variable.
The truth values $\mathsf{Bool} \triangleq \{\texttt{true},\texttt{false}\}$. are ranged over by a metavariable $t$, and $n$ is the metavariable for $\mathbb{Z} \triangleq \{\cdots, -1, 0, 1, 2, \cdots\}$.
The metavariable $d \in \mathbb{D}$ ranges over data values, including 8-, 32-, and 64-bit integers as well as single- and double-precision floating-points.

Control expressions $\mathrm{CExpr}$ encompass integer, scalar read, and (quasi-) affine arithmetic operations.
Boolean expressions $\mathrm{BExpr}$ define basic logical and relational operations on control expressions.
Data expressions $\mathrm{DExpr}$ are similarly defined, but operations need not be affine.
Barrier Expressions $\mathrm{ZExpr}$ are only allowed to read, but not composed.
Window coordinates $w^*$ (we use $.^*$ to denote a tuple) used in data and barrier read accesses are specified as a tuple of point-wise or interval control expressions, expressing a windowed access to multi-dimensional arrays.
Scalar read is syntactic sugar for an array access of $w^*=\langle\,\rangle$ (empty tuple).


\vspace{-0.7em}
\subsection{Types}
\label{sec:gpuir-types}



All types are fully defined in Appendix~\ref{sec:gpuir-type-figures}. This section provides a general introduction to the different types in \ir{}.
Figure~\ref{fig:variable-types} defines data and barrier types.
Data types $\tau_x$ specify precisions for data variables $\mathbb{X}$, such as \texttt{f32} and \texttt{i32}.
Barrier types $\tau_z$ distinguish a non-explicitly guarded barrier from an explicitly guarded barrier (\texttt{barrier(z)}); the explicitly guarded barrier case is required only for mbarriers.%
\footnote{This requirement has been removed from newer Exo-GPU versions.}
%
A \emph{collective type} ($\delta \in \Delta$) defines a tuple consisting of a domain and a box, each representing $d$- dimensional natural numbers ($(\mathbb{N}^d, \mathbb{N}^d)$).
Together, these define a $d$-level thread hieararchy within a thread cluster, and a selection of a number of elements on each level.
%
An argument to \texttt{cuda\_threads} loop $\tau_u$ is parameterized by a collective type $\delta$, which is defined in Figure~\ref{fig:CollUnit}.
%
Figures~\ref{fig:CudaMemory}–\ref{fig:SpecialWindow} classify three categories of memory spaces: $\pi_\mathrm{d}$ for data memories (such as \texttt{CudaRmem} for CUDA registers), $\pi_\mathrm{z}$ for barrier completion mechanisms (such as \texttt{mbarrier} or \texttt{commit\_group}), and $\pi_\mathrm{w}$ for special window aliasing (such as \texttt{CUtensorMap}).
Similar to $\tau_u$, $\pi_\mathrm{d}$ is parameterized by $\delta$.

A qualitative timeline $\tau_q$ annotates each \emph{runtime} buffer access.
We define distinct $\tau_q$ to distinguish between buffer accesses that require different synchronization mechanisms or memory fences to resolve hazards.
For example, we distinguish non-async copies from \texttt{cp.async} copies (which require \texttt{cp.async.wait\_all} or similar to wait for), and we distinguish register and SMEM operands of \texttt{wgmma} instructions (which require \texttt{wgmma.fence} and \texttt{fence.proxy.async}, respectively).
Section~\ref{sec:abstract-machine} will formally define how a runtime buffer access instance $x[i_1,\dots,i_n]$ tagged with $q_1$ states that, at the program point and thread, the access to $x$ occurs under $q_1$.
%
Moreover, the sync timelines $\mathrm{SyncTL}$ mentioned in Section~\ref{sec:language-overview}, which parameterize synchronization statements, are actually defined as compositions of qualitative timelines ($\mathrm{QualTL}$), as shown in Figure~\ref{fig:SyncTL-def}.

\vspace{-0.7em}
\subsection{Programs}
\label{sec:gpuir-progams}

\input{tex_gpu/grammar_gpu_ir_stmts}

\name{} supports two types of procedures, $p: \Proc$, which models CPU functions (possibly containing CUDA kernel launches), and $g: \mathrm{Instr}$, which models hardware instructions for either CPU or CUDA.
Both procedure types share three common components: a statement body that defines sequential behavior, a tuple of control parameters $y^*$, and a tuple of data parameters $x^*$.
Each data parameter in $\Proc$ carries two annotations: $\tau_x$, specifying the required data precision, and $\pi_d$, specifying the required memory type for the passed argument.

Instruction parameters ($\tau_a$) also carry both $\tau_x$ and $\pi_d$ annotations, as well as hardware-specific information, namely: out-of-order, convergence, and qualitative timeline information, which all affect only abstract machine semantics for checking (Section~\ref{sec:abstract-machine}).
The $\tau_a$ also contains a distributed collective units tuple ($\tau_u^*$) specifying the CUDA thread hierarchy at which each array dimension is sharded (checked by distributed memory analysis,~Section~\ref{sec:distributed-memory-analysis}).
%
The instruction ($g$) contains annotations for $t$ (a CPU/CUDA flag), a collective unit $\tau_u$ specifying the convergent threads required, and $\pi_z$ and $\tau_u^*$. 
%


\vspace{-0.7em}
\subsection{Statements}
\label{sec:gpuir-stmts}

Figure~\ref{fig:gpuir-statement-syntax-def} defines statements $s\in\Stmt$.
A statement can sequence ($s_1; s_2$), iterate over loops (\texttt{seq}, \texttt{cuda\_tasks}, or \texttt{cuda\_threads}), and guard (\texttt{if} $b$ \texttt{then} $s$).
Data effects include indexed writes $x[c^*]=e$, reductions $x[c^*]{+}{=}\,e$, and window creation $x=x[c_{\mathrm{lo}}:c_{\mathrm{hi}}]\ @\ \pi_{\mathrm{w}}$.
\texttt{CudaDeviceFunction} and \texttt{CudaWarps} control generation and selection of CUDA threads.
Synchronization statements \texttt{Fence}, \texttt{Arrive}, and \texttt{Await} get lowered to CUDA synchronization and proxy fence instructions.



\begin{wrapfigure}{r}{0.49\textwidth}
\vspace{-5mm}
\begin{minted}[fontsize=\footnotesize,breaklines]{python}
with CudaDeviceFunction(...):
    for cta_m in cuda_tasks(...):
        for cta_n in cuda_tasks(...):
            # Device task starts here
            A_smem: f32[128, 32] @ CudaSmemLinear
            # ...
\end{minted}
\vspace{-8mm}
\end{wrapfigure}
\textbf{CUDA Kernel Structure:}
All \name{} code is compiled as CPU code by default.
A \texttt{CudaDeviceFunction} block specifies a CPU-side launch of a CUDA kernel.
Its body must contain only a single statement: a nest of one or more \texttt{cuda\_tasks} loops; \texttt{cuda\_tasks} loops must not appear elsewhere.
Each iteration of the inner-most \texttt{cuda\_tasks} loop comprises a device task and is assigned to one CUDA cluster for execution.
The \emph{device scope} of a statement is CPU if outside of a \texttt{CudaDeviceFunction} block, and CUDA if inside.
At CPU scope, all loops must have loop mode \texttt{seq}.
The arguments of \texttt{CudaDeviceFunction($n, e_w^*$)} respectively define the \texttt{clusterDim} and, indirectly, the \texttt{blockDim} of the CUDA clusters launched.
Each warp expression $e_w: r = (n, n, n)$ defines a group of warps named $r$, with the first $n$ parameter being the number of warps.
If nonzero, the latter two $n$ parameters specify that the warps will adjust the register count down or up, respectively, using a \texttt{setmaxnreg.dec} or \texttt{setmaxnreg.inc} instruction \cite{ptx}.
We infer that \texttt{blockDim} is 32 times the total number of warps named.


\begin{wrapfigure}{r}{0.5\textwidth}
\vspace{-2mm}
\begin{minted}[fontsize=\footnotesize,breaklines]{python}
my_warp_config = [
  CudaWarpConfig("producer", 1, 40, 0),
  CudaWarpConfig("unused",   3, 40, 0),
  CudaWarpConfig("consumer", 8, 0, 232)]
with CudaDeviceFunction(clusterDim=2,
    warp_config=my_warp_config):
\end{minted}
\vspace{-5mm}
\caption{Launching clusters of 2 CTAs each, with 1 ``producer'', 3 ``unused'', and 8 ``consumer'' warps per CTA (total \texttt{blockDim=384}). The consumer warps have 232 registers per thread, and the others only 40.}
\label{fig:warp-config}
\vspace{-2mm}
\end{wrapfigure}
\textbf{Thread Control:}
Having described the \texttt{cuda\_tasks} loops (inter-cluster parallelism), we now describe \texttt{cuda\_threads} loops and \texttt{with\;CudaWarps} (intra-cluster parallelism).
A \texttt{cuda\_threads} loop takes its executing thread collective and sub-divides, assigning one subdivision of threads to execute each loop iteration (possibly with left-over, inactivated threads).
The \texttt{unit} parameter adjusts the number of threads per iteration.
The \texttt{CudaWarps($n,n,r$)} block restricts its body statement to execute only with threads residing in the warp variable named by $r$ (which is defined in the \texttt{CudaDeviceFunction}); the numeric arguments further restrict the warps selected.


\textbf{Proc \& Instruction Calls:}
Calls to $p: \Proc$ and $g: \mathrm{Instr}$ are syntactically similar, distinguished only by the callable's type.
\name{}'s static analysis enforces that calls to $p: \Proc$ only appear at CPU scope, while calls to $g: \mathrm{Instr}$ appear at CPU or CUDA scope as appropriate for $g$.
For CUDA instructions, calls to $g$ must only appear at $\tau_u$-scope, where $\tau_u$ is the collective unit specified for $g$.
Additionally, each data parameter must have the correct precision $\tau_x$ and memory type $\pi_d$, and the trailing barrier (if required) must have the correct $\pi_z$.
%

\textbf{Synchronization Statements:}
%
The user specifies synchronization between threads and/or between async accelerator instructions by inserting synchronization statements.
They take \emph{synchronization timeline} ($\mathrm{SyncTL}$) parameters, defined as compositions of qualitative timelines and a transitivity (trnstv?) flag (Figure~\ref{fig:SyncTL-def}).
The transitivity flag controls interaction between synchronization statements, detailed in Section~\ref{sec:abstract-machine}.

Each $\tau_s:\mathrm{SyncTL}$ contains two \textsf{QualTL} sets: the full timeline set (indicated by ``full'' in (Figure~\ref{fig:SyncTL-def}), and a temporal timeline set (indicated by ``full'' or ``temp.'').
A \texttt{Fence(}$\tau_s^\text{pre}, \tau_s^\text{post}$\texttt{)}, or paired \texttt{Arrive(}$\tau_s^\text{pre}, $\texttt{\_)} and \texttt{Await(\_}, $\tau_s^\text{post}, $\texttt{\_)} statements, resolves hazards between two accesses to the same array element when: (1) the first access precedes synchronization and uses a $\mathrm{QualTL}$ from $\tau_s^\text{pre}$'s full timeline set, and (2) the second access follows synchronization and uses a $\mathrm{QualTL}$ from $\tau_s^\text{post}$'s full timeline set (for reads) or temporal timeline set (for writes).
In hardware terms, the distinction between the full and temporal timeline sets is due to memory fences.
If the prior value of a buffer element is \emph{overwritten}, without being read, we require only that the overwrite is temporally ordered after prior accesses to that buffer element, and memory fences may be elided.




\textbf{Alloc \& Free:}
An allocation statement begins the lifetime of a new data or barrier variable.
For data allocations, the data memory $\pi_d$ determines the physical backing memory (registers (RMEM), shared memory (SMEM), or global memory (GMEM)) and the mapping between multidimensional tensor coordinates and 1D memory offsets (Figure~\ref{fig:CudaMemory}).
It also carries an implicit collective type $\delta$, specifying which level of the thread hierarchy at which the physical memory type is allocated (e.g. CTA for shared memory allocations); this is consumed by distributed memory analysis (Section~\ref{sec:distributed-memory-analysis}).
For barrier allocations, the $\mathrm{BarrierComp}$ parameter $\pi_z$ controls the completion mechanism used for CUDA code translated from \texttt{Arrive} and \texttt{Await} statements (Figure~\ref{fig:BarrierType}).
The explicitly-guarded barrier type is relevant only for certain configurations of \texttt{mbarrier} usage.
A free statement ends the lifetime of a data or barrier variable.
Currently, we forbid the user from inserting free statements manually; these are added automatically in a separate compiler pass.
%


\textbf{Window Statement:}
A window statement \texttt{$x_\text{win}$ = $x_\text{data}$[$w^*$] @ $\pi_w$} defines an alias to $x_\text{data}$.
A window into $x_\text{win}$ can be passed as an instruction argument which requires special window type $\pi_w$.
Currently, we only use this feature to construct \texttt{CUtensorMap} objects~\cite{ptx} required for TMA instructions; we parameterize the \texttt{CUtensorMap} type with its swizzle mode and box shape.

%% file: tex_gpu/grammar_gpu_ir_exprs.tex
\begin{figure*}[t]
\footnotesize
\centering
\setlength{\tabcolsep}{2pt}
\arraycolsep=1.8pt\def\arraystretch{1.0}

\begin{minipage}[t]{0.48\textwidth}
\centering
\begin{tabular}{rrll}
\toprule
$v: \textsf{Var}$ & $\Coloneqq$ &
      $x \in \mathbb{X} $ & data variable\\
      &|& $y \in \mathbb{Y} $ & control variable\\
      &|& $z \in \mathbb{B} $ & barrier variable\\
      &|& $r \in \mathbb{W} $ & warp variable\\
\midrule
$e : \mathrm{DExpr}$ & $\Coloneqq$ &
      $d$ & data value \\
  &|& $x$ | $x\texttt{[}w^*\texttt{]}$ & reads \\
  &|& $e_1{+}e_2$ | $e_1{-}e_2$ | $e_1{*}e_2$ | $e_1{/}e_2$ & compound expr \\[3pt]
$e_\mathrm{z} : \mathrm{ZExpr}$ & $\Coloneqq$ &
  $z$ | $z\texttt{[}w^*\texttt{]}$ & reads \\[3pt]
$e_\mathrm{w} : \mathrm{WExpr}$ & $\Coloneqq$ &
  $r~~\texttt{=}~~(n, n, n)$ & warp config \\
\midrule
$n$ & $\in$ & $\mathbb{Z}$ & integer \\
$t$ & $\in$ & $\mathsf{Bool}$ & boolean \\
$d$ & $\in$ & $\mathbb{D}$ & data values \\
\bottomrule
\end{tabular}
\end{minipage}
\hfill
\begin{minipage}[t]{0.48\textwidth}
\centering
\begin{tabular}{rrll}
\toprule
$c : \mathrm{CExpr}$ & $\Coloneqq$ &
      $n$ & integer \\
      &|& $y$ & reads \\
      &|& $c_1{+}c_2$ | $c_1{-}c_2$ | $c{*}n$ & affine arith \\
      &|& $c~{/}~n$ | $c~\texttt{mod}~n$ & quasi affine \\[3pt]
$b : \mathrm{BExpr}$ & $\Coloneqq$ &
  $t$ & boolean \\
  &|& $b_1\,\texttt{and}\,b_2$ | $b_1\,\texttt{or}\,b_2$ & logical ops \\
  &|& $c_1\,{==}\,c_2$ | $c_1{<}c_2$ | $c_1{\leq}c_2$ & relational ops \\[3pt]
$w : \mathrm{WinCoord}$ & $\Coloneqq$ &
     $c$ & point access \\
     &|& $c_1..c_2$ & interval access\\
\bottomrule
\end{tabular}
\end{minipage}
\vspace{-0.5em}
\caption{The syntax of variables and expressions of \ir{}.
We use $\cdot^*$ to mean 0 or more.}
\label{fig:gpuir-expression-syntax-def}
\vspace{-1.5em}
\end{figure*}

%% file: tex_gpu/grammar_gpu_ir_stmts.tex
\begin{figure}[t]
\footnotesize
\centering
\arraycolsep=1.8pt\def\arraystretch{1.0}
\setlength{\tabcolsep}{2pt}
\begin{tabular}{rrll}
\toprule
  $s : \Stmt$ & $\Coloneqq$ &
        $s_1 \texttt{;} s_2$ & sequencing \\
 &$|$& \texttt{for $y$ in \texttt{seq}($c_{\mathrm{lo}}$,$c_{\mathrm{hi}}$) do $s$} & sequential loop \\
 &$|$& \texttt{for $y$ in \texttt{cuda\_tasks}($c_{\mathrm{lo}}$,$c_{\mathrm{hi}}$) do $s$} & cuda tasks \\
 &$|$& \texttt{for $y$ in \texttt{cuda\_threads}($c_{\mathrm{lo}}$,$c_{\mathrm{hi}}$,\texttt{unit=}$\tau_u$) do $s$} & cuda threads \\
  &$|$& \texttt{with CudaDeviceFunction(}$n,~e_\mathrm{w}^*$\texttt{) do s} & CUDA device function block \\
  &$|$& \texttt{with CudaWarps($n, n, r$) do s} & warp specialization block \\
  &$|$& \texttt{if $b$ then $s$} & guard \\
  &$|$& $p(c^*, e^*)$ & non-instruction subprocedure call \\
  &$|$& $g(c^*, e^*)$  \; \texttt{>}\texttt{>} \; $e_\mathrm{z}$ & instruction subprocedure call \\
  &$|$& $x\texttt{[}c^*\texttt{] = } e$ & write \\
  &$|$& $x\texttt{[}c^*\texttt{] += } e$ & reduce \\
  &$|$& $x = x[w^*] \, @ \, \pi_\mathrm{w}$ & window statement\\
  &$|$& \texttt{$x:\tau_x[c^*] \, @ \, \pi_\mathrm{d}$} & data alloc \\
  &$|$& \texttt{$z:\tau_z[c^*] \, @ \, \pi_\mathrm{z}$} & barrier alloc \\
  &$|$& \texttt{free $x$} & data free \\
  &$|$& \texttt{free $z$} & barrier free \\
  &$|$& \texttt{Fence(}$\tau_s$, $\tau_s$\texttt{)} & non-split barriers \\
  &$|$& \texttt{Arrive(}$\tau_s$, $n$\texttt{)} \; \texttt{>}\texttt{>} \; $e_\mathrm{z}^*$ & arrive \\
  &$|$& \texttt{Await(}$e_\mathrm{z}$, $\tau_s$, $n$\texttt{)} & await \\
\midrule
$\tau_a : \mathrm{Arg}$ & $\Coloneqq$ &
    $\tau_x$ & data type (precision) \\
    &$|$& $\pi_\mathrm{d}$ & data memory type \\
    &$|$& $\texttt{out\_of\_order} \in \mathsf{Bool}$ & out-of-order execution flag \\
    &$|$& $\texttt{convergent} \in \mathsf{Bool}$ & implicit thread sync flag \\
    &$|$& $\tau_u^*$ & distributed collective units \\
    &$|$& $\texttt{initial\_qual\_tl} \in \mathrm{QualTL}$ & initial qualitative timeline \\
    &$|$& $\texttt{ext\_qual\_tl} \in \mathcal{P}(\mathrm{QualTL})$ & extended qualitative timeline set \\
    &$|$& $\texttt{atomic\_qual\_tl} \in \mathcal{P}(\mathrm{QualTL})$ & atomic qualitative timeline set \\
\midrule
\multicolumn{4}{l}{
\hspace{1mm} $p : \Proc$ $\Coloneqq$
        $\begin{array}{l}
          \texttt{proc}~~ y^*, (x:\tau_x \texttt{\;@\;} \pi_d)^* \\
          ~~\texttt{do}~s \\
        \end{array}$
\hspace{1cm}
$g : \mathrm{Instr}$ $\Coloneqq$
        $\begin{array}{l}
          \texttt{proc}~~ y^*, (x: \tau_a)^*, t, \tau_u, \pi_\mathrm{z}, \tau_u^* \\
          ~~\texttt{do}~s \\
        \end{array}$
}\\
\bottomrule
\end{tabular}
\vspace{-0.5em}
\caption{The syntax of \ir{} statements.
}
\vspace{-2.5em}
\label{fig:gpuir-statement-syntax-def}
\end{figure}

%% file: tex_gpu/03_collective_analysis.tex
\section{Collective Analysis}
\label{sec:collective-analysis}

\newcommand{\derivedCollTiling}{\mathsf{derive}_\omega}

\newcommand{\tileCount}{\mathsf{tileCount}}

\newcommand{\N}{\mathbb{N}}
\newcommand{\Y}{\mathbb{Y}}


Collective analysis is a static, forward dataflow analysis over a GPU IR procedure. Its results are consumed by
(1) a memory analysis, (2) code generation, and (3) an abstract‑machine interpreter.  
The analysis annotates every GPU lexical scope with a collective tiling $\omega \in \Omega$. 
By default, statements inherit their parent’s tiling unchanged; a new tiling is derived only at: device entry (\texttt{CudaDeviceFunction}), thread loops (\texttt{cuda\_threads}), and warp‑specialization blocks (\texttt{CudaWarps}).
The purpose of this analysis is to statically reason about the mapping between work and the threads \emph{within} clusters, so here we ignore \texttt{cuda\_tasks} loops, which distribute work across clusters.
We uniquely identify a thread within a cluster by its \emph{natural thread index}; a set of such indices comprises a \emph{thread collective} $\mu \in \mathbb{T}$, where $\mathbb{T} \;\triangleq\; \mathcal{P}(\mathbb{N})$.

\begin{definition}[Natural thread index]
The \emph{natural thread index} for a CUDA thread is \texttt{cluster\_ctarank * blockDim.x + threadIdx.x}. (\name{} only parallelizes on the x dimension).
\end{definition}

\begin{definition}[Collective tiling]
A \emph{collective tiling} $\omega \in \Omega$ for an $M$-dimensional domain is an $M$‑tuple of \emph{dimension descriptors}
\[
\omega = \langle \mathcal{D}_0,\ldots,\mathcal{D}_{M-1}\rangle
\qquad\text{with}\qquad
\mathcal{D}_m = (D_m,\, \mathcal{O}_m),
\]
where $D_m \in \N$ is the extent of dimension $m$ and $\mathcal{O}_m \in \mathcal{O}^*$ is an ordered list of (possibly empty) \emph{dimension operators} $\mathcal{O}: \Y \times \N^3$.
We write the thread pitch of dimension $m$ as
\[
P_m \;\triangleq\; \prod_{i=m+1}^{M-1} D_i
\;\text{ where } P_{M-1}=1
\]
\end{definition}

The dimensions of a collective tiling $\omega$ capture how a cluster's threads are partitioned into hierarchical collectives (cluster, block, warp).
We support \emph{reshape} to introduce ad-hoc additional hierarchy (e.g. pairs of warps).
From there, each dimension's dimension operators summarize the control variables (identified by $y \in \mathbb{Y}$) that iterate on that dimension.
The thread pitch of the dimension describes the distance, in units of natural thread indices, between adjacent elements (e.g. a CTA dimension would have thread pitch \texttt{blockDim}).

During forward dataflow analysis, new collective tilings are derived for the three special statements mentioned above, while all other statements inherit their parent's tiling.
It produces a child collective tiling through the following transformation:
\[
\derivedCollTiling:\;(\omega^{\mathrm{env}},\delta^{\mathrm{stmt}}, y,\mathsf{lo},\mathsf{hi},\tileCount)\;\rightarrow\;\omega'
\]
%
where $\omega^{\mathrm{env}}$ is the parent scope’s collective tiling and $\delta^{\mathrm{stmt}}$ is the collective type from the statement.
The (optional) iterator $y$ and bounds $[\mathsf{lo},\mathsf{hi})$ describe a regular slice of linear thread space; $\tileCount\in\N$ is the number of tiles for that slice.
The details of $\derivedCollTiling$ are left for future work.

The \texttt{cuda\_threads} loop and \texttt{CudaWarps} block use $\derivedCollTiling$ to create a new collective tiling $\omega$ based on the one annotating the parent scope.
The \texttt{CudaDeviceFunction} block derives $\omega$ based on launch parameters: $\omega$ is 1‑D if \texttt{clusterDim}$=1$ (domain $D_0{=}\texttt{blockDim}$), otherwise 2‑D with $(D_0,D_1){=}(\texttt{clusterDim},\texttt{blockDim})$; all operator lists start empty.


The implementation of $\derivedCollTiling$ proceeds in two steps:
\begin{enumerate}
  \item \emph{Shape alignment.} Make $\omega^{\mathrm{env}}$ and $\delta^{\mathrm{stmt}}$ compatible by reshaping via hierarchical splits so that their thread pitches $P_m$ align.
  \item \emph{Single‑dimension refinement.} After alignment, at most one dimension $m$ is newly tiled; the derivation appends exactly one operator to $\mathcal{O}_m$.
\end{enumerate}

Once a tiling $\omega$ is derived, it is compiled into a \emph{thread mapping}:
$f_\omega : \Sigma \to \mathbb{T}$, where $\Sigma \triangleq (\mathbb{Y} \to \mathbb{Z})$
which maps a control environment $\sigma \in \Sigma$ to a runtime thread collective $\mu \in \mathbb{T}$.  Equivalently, let
$\mathcal{F} \triangleq \{\, f \mid f : \Sigma \to \mathbb{T}\,\}$
and
$f_\omega \in \mathcal{F}$.
During conversion, each Abstract Machine statement $s^\#$ is annotated with its thread mapping $f_{s^\#} \in \mathcal{F}$.  In Section~\ref{sec:amir-semantics} we write $f_{s^\#}$ for the thread mapping associated with statement $s^\#$.

\subsection{Distributed Memory Analysis}
\label{sec:distributed-memory-analysis}



\vspace{-0.3em}
\emph{Distributed memory analysis} is a static, program-wide consistency check that ensures every use of CUDA-scoped data and barrier variables is compatible with their memory definitions.
Each memory type $\pi_d$ specifies a collective type $\delta_{\pi_d}$.
The analysis validates that  allocation and use agree on which threads ``own'' which memory slices, indexing respects that ownership, and all uses induce the same mapping from logical indices to CUDA thread identifiers.
Analysis runs in four steps:
\vspace{-0.2em}
\begin{enumerate}
  \item \textbf{Desugaring instruction call.} 
  To expose the implicit collective requirement encoded by each hardware instruction to the collective analysis in \textbf{step~2}, we rewrite instruction calls $g(\cdots)$ into explicit loops. Each arguments to the instruction call are wrapped around with \texttt{cuda\_threads}, with their collective units specified by $g.\tau_u^*$ in the instruction specification.
  \item \textbf{Collective analysis.} Run collective analysis and annotate each scope with $\omega\in\Omega$.
  \item \textbf{Triple collection.} For each data or barrier variable, record the \emph{access triple} of all of use sites. We use $\mathcal{T}_v$ to denote a set of those triples for a variable $v$; these sets for all the data and barrier variables are the sole inputs to \textbf{step~4}. 
  Each triple $(n^*,\omega,c^*)$ denotes: the subdivided dimensions $n^*$, the use-site tiling $\omega$, and observed index coordinates $c^*$. For each use site:
    \begin{itemize}
      \item \textbf{Data.} Compute $n^*$ from the memory's collective type $\delta_{\pi_d}$ and the use-site tiling $\omega$. 
      \item \textbf{Barriers.} Treat all dimensions as subdivided (every axis participates in distribution).
    \end{itemize}
  \item \textbf{Triple analysis.} 
    For each CUDA-scoped data variable $v$, we require: (i) consistency of the allocation w.r.t. the memory's $\delta_{\pi_d}$, (ii) consistency of data or barrier accesses across sites (by checking all triples in $\mathcal{T}_v$), and (iii) all triples in $\mathcal{T}_v$ must derive one and only one thread pitch tuple that summarizes how incrementing logical indices advances natural thread IDs.
    The derivation will be detailed in future work.
\end{enumerate}
\vspace{-0.2em}
\noindent
As a result, distributed memory analysis rejects programs that (i) mismatch allocation and accesses with their memory declaration, (ii) inconsistently index the same data or barrier across sites, or (iii) rely on missing/extra CUDA-thread iterators. Passing guarantees that the program’s memory behavior is coherent with its collective structure and declared memory types.

\vspace{-0.7em}
\subsection{Code Generation to CUDA Code}
\vspace{-0.2em}


\begin{wrapfigure}{r}{0.5\textwidth}
\vspace{-5mm}
\begin{minted}[fontsize=\footnotesize,breaklines]{python}
if (int y = f(blockIdx, threadIdx); g(y)) {
    lower(s);}
\end{minted}
\vspace{-6mm}
\end{wrapfigure}
\textbf{Lowering \texttt{cuda\_threads} Loops:}
The correspondence between parallel loops and threads has thus far been expressed in terms of thread mappings of type $\Sigma \to \mathbb{T}$, which infer a set of active threads from the control environment $\sigma$.
However, the generated CUDA C++ implements the inverse of this mapping, inferring the value of a control variable \texttt{y} from the thread index instead.
Each \name{} loop of the form \texttt{for y in cuda\_threads(lo, hi, unit=$\tau_u$) do s} lowers to CUDA C++ as shown in the inset right. Here, \texttt{f} is a function of CUDA's built-in \texttt{blockIdx} and \texttt{threadIdx} variables, and \texttt{g} is a boolean condition on \texttt{y} that disables threads not assigned to execute any iteration of the body.
Where possible, \texttt{f} is 0 and \texttt{g} is 1, to facilitate uniform branching.
%

\textbf{Warp Specialization:}
When a device function uses multiple warp variables $r_1, ..., r_W \in \mathbb{W}$, we generate $W$-many CUDA C++ code paths from a single \name{} device function, each specialized for (and executed by) the threads assigned to one warp variable.
Each path uses PTX's \texttt{setmaxnreg} instruction to vary per-thread register counts, and omits all code paths guarded by a \texttt{with} \texttt{CudaWarps} block holding code not intended for the current warp.
This allows the PTX assembler to statically verify that register-heavy instructions (e.g. \texttt{wgmma}) don't appear on low-register code paths.

\textbf{Persistent Kernels and \texttt{cuda\_tasks} Loops:}
The generated CUDA C++ device functions implement a persistent kernel design that launches exactly the number of thread block clusters needed to saturate the device.
Following the approach of CUTLASS grouped kernel schedulers~\cite{grouped_kernel_schedules}, we implement a C++ task generator object that generates a series of device task coordinates for thread block clusters to execute.
The current implementation orders tasks lexicographically and assigns them round-robin to thread block clusters for execution.
While we leave it as future work to provide a mechanism for programmers to customize this ordering, such customization would be unconstrained by frontend's quasi-affine indexing restrictions since \texttt{cuda\_tasks} loops impose no semantic ordering, thereby enabling techniques like Morton swizzling.


\textbf{Shared Memory Allocations:}
Device functions generated by \name{} use only dynamic shared memory for SMEM allocation.
The \name{} compiler deduces the required SMEM allocation size and statically assigns offsets into this allocation for each SMEM-backed variable.
Variables with non-overlapping lifetimes may share the same memory locations; to ensure safe aliasing, we prohibit any thread in the cluster from passing an SMEM free statement until all reads and writes to the freed variable have retired, as enforced by the Abstract Machine (Section~\ref{sec:abstract-machine}).
This is necessary because freed SMEM may be immediately reused by any thread in the cluster.
We note that the compiler's aliasing may be suboptimal, and the cluster-wide safety requirement may be overly conservative (e.g., for CTA-local variables). Providing programmers explicit control over SMEM lifetime and aliasing is left as future work.



%% file: tex_gpu/02_abstract_machine.tex
\vspace{-0.5em}
\section{Abstract Machine}
\label{sec:abstract-machine}
\vspace{-0.1em}

To guarantee sequential–parallel equivalence, \name{} performs synchronization checking  via Abstract Machine interpretation.
This section introduces the AMIR grammar (Section~\ref{sec:amir-grammar}), the translation from \ir{} (Section~\ref{sec:conversion}), and the AMIR execution semantics (Section~\ref{sec:amir-semantics}).
We translate \ir{} into AMIR solely for validation: the frontend language remains simple, while synchronization reasoning is factored into explicit AMIR operations that are orthogonal to value computation.

\vspace{-0.5em}
\subsection{Abstract Machine Grammar}
\label{sec:amir-grammar}
\vspace{-0.1em}

\input{tex_gpu/grammar_abs_machine}

Figure~\ref{fig:am-syntax-def} defines AMIR grammar.
It follows \ir{} for control $c$, booleans $b$, window coordinates $w$, and all variable/type definitions (Figures ~\ref{fig:gpuir-expression-syntax-def},~\ref{fig:gpuir-statement-syntax-def}).
AMIR defines only windowed access expressions, written $e_d^\# \triangleq x[w^*]$ (data access) and $e_\mathrm{z}^\# \triangleq z[w^*]$ (barrier access).
%
AMIR introduces visibility levels $\tau_v \in \mathrm{VL}$, which are totally ordered:
$\texttt{invisible} \prec \texttt{atomic\_only} \prec \texttt{unordered} \prec \texttt{temporally\_ordered} \prec \texttt{fully\_ordered}$. 

AMIR has no hardware-instruction procedures ($g$ in \ir{}) and no subprocedure calls.
Procedures $p^\#$ take control-only arguments.
The statement (Figure~\ref{fig:am-syntax-def}) consists of:
(i) structural statements--sequencing, loop, and guard
(ii) recording statements that update the synchronization environment \textsf{SyncEnv}--\texttt{Alloc}, \texttt{RecordRead}, \texttt{RecordMutate}, \texttt{Fence}, \texttt{Arrive}, and \texttt{Await};
and (iii) checking statements that query \textsf{SyncEnv} and fail if the required visibility obligations are not met; otherwise they are no-ops: \texttt{CheckBarrier}, \texttt{CheckReads}, and  \texttt{CheckMutates}.

\vspace{-0.7em}
\subsection{Conversion from GPU IR to Abstract Machine IR}
\label{sec:conversion}

\input{tex_gpu/convresion}
\vspace{-0.1em}

We describe how \ir{} expressions (Section~\ref{sec:expression-conversion}), statements (Section~\ref{sec:statement-conversion}), and instruction arguments (Section~\ref{sec:instr-arg-conversion}) are converted.
%
Before conversion, we apply two trivial preprocessing steps to \ir{}: (i) Inline all window statements, and (ii) Inline all sub\-procedure calls that occur outside any \texttt{CudaDeviceFunction} block.
%
Throughout this section (and in figures) we use the following notation.
For a finite, ordered variables $R=\langle r_0,\dots,r_{k-1}\rangle$ and a statement template $S(\cdot)$,
$\bigseq_{i=0}^k S(x)
\;\triangleq\;
S(r_0)\;;\;S(r_1)\;;\;\cdots\;;\;S(r_{k-1})$.
If $R=\emptyset$, the result is no‑op.
The intent is purely left‑to‑right sequential composition at AMIR conversion time.
When unambiguous, we write $f(A)=\{\,f(a)\mid a\in A\,\}$ for the pointwise image of a set.





\vspace{-0.2em}
\subsubsection{Expression Conversion}
\label{sec:expression-conversion}

Figure~\ref{fig:expression_conversion} defines
$T_e : \mathrm{DExpr}\cup\mathrm{ZExpr}\to\mathcal{P}(\Expr^\#)$,
which extracts the windowed AMIR accesses $e^\#\in\{x[w^*],\,z[w^*]\}$:
scalars $d$ contribute nothing ($\emptyset$); a bare name $x$ or $z$ yields the degenerate accesses $x[\langle\rangle]$ or $z[\langle\rangle]$;
explicit indexings $x[w^*]$ and $z[w^*]$ are preserved; and for arithmetic $e_1~\mathrm{op}~e_2$ the sets union, $T_e(e_1)\cup T_e(e_2)$.
$T_e$ is set-valued (order and duplicates are irrelevant), leaves the value expression unchanged, and merely supplies the converted $\Expr^\#$ used by \texttt{RecordRead}/\texttt{RecordMutate} and subsequent visibility checks.

\vspace{-0.2em}
\subsubsection{Statement Conversion}
\label{sec:statement-conversion}

Figure~\ref{fig:statement_conversion} gives a structural translation
$T_s : \mathrm{Stmt} \to \mathrm{Stmt}^\#$ from GPU IR to AMIR.
Helper functions referenced by Figure~\ref{fig:statement_conversion} are defined in Figure~\ref{fig:statement_conversion_helpers}.
Sequencing and conditionals are preserved, while all loops are normalized to the abstract \texttt{loop} form.
Warp/block scopes are dropped and replaced with a guard whose condition is \texttt{true}.
Assignments $x[c^*] \mathrel{\diamond} e$ ($\diamond \in \{=,\mathrel{+}=\}$) are lowered to explicit read/write effects.
We expand source and destination with $T_e$ and remove sync‑exempt expressions via $T_\mathrm{ex}$.
\emph{Sync‑exempt memories} are those whose accesses can be assumed safe without checking:
\texttt{CudaGridConstant}, \texttt{CudaClusterSync}, and \texttt{CudaCommitGroup}.
\emph{Shared memories} are the following:
\texttt{CudaBasicSmem}, \texttt{CudaSmemAtomicity16B}, \texttt{CudaSmemLinear}, \texttt{Sm90\_SmemSwizzled(B)}, or \texttt{CudaMbarrier}.

Hardware instruction calls ignore control arguments and pair the instruction's parameter definition and the arguments via $T_{\mathrm{arg}}$.
The entry and exit of \texttt{CudaDeviceFunction} is bracketed with fences with \texttt{cuda\_stream\_sync}, capturing both the CPU-to-GPU fence implied by the kernel launch, and the serialization of the CUDA kernel with respect to prior and subsequent CUDA kernels and API calls (Note that \name{} currently uses only one CUDA stream).
During the conversion, we record the scope of each statement using a global environment $\mathsf{Scope} : \mathrm{Stmt}^\# \to \{ \text{CPU}, \text{CUDA} \}$. Statements outside the \texttt{CudaDeviceFunction} block are CPU-scoped, while those inside are GPU-scoped. The $\mathrm{scope}$ mapping is used in Definition~\ref{def:global-threads-collective}.
\texttt{Fence}, \texttt{Arrive}, and \texttt{Await} translate a sync timeline ($\tau_s$) into a qualitative timeline set ($\tau_q^*$).
$T_\mathrm{tr}$ queries whether a sync timeline is transitive; $T_\mathrm{qfull}$ and $T_\mathrm{qtmp}$ materialize the full and temporal $\mathrm{QualTL}$ sets, respectively.
\texttt{Arrive} selects its arrival qualitative timeline via $T_\mathrm{qarr}$ to sync-check the barrier itself, thereby preventing use-after-free bugs.









\vspace{-0.2em}
\subsubsection{Instruction Argument Conversion}
\label{sec:instr-arg-conversion}


Figure~\ref{fig:targ_conversion} expands the conversion $T_\mathrm{arg}(g.p_i, e_i, e_z)$ from a \ir{} instruction call argument to the AMIR statements.
The auxiliary projections and predicates referenced in
the figure---$T_\mathrm{ext}$, $T_\mathrm{atom}$, $T_\mathrm{init}$,
$T_\mathrm{ooo}$, and $T_\mathrm{cvg}$---are defined in
Figure~\ref{fig:targ_conversion_helpers}.
For each non-sync exempt expression $e^\# \in T_\mathrm{ex}(e_i)$, the conversion emits a sequence of
checks and effect-recording operations which depends on two
properties of the parameter $g.p_i$: its \emph{access mode}
(read-only vs.\ write-only/read--write) and whether it is
\emph{atomic}.
The access mode of $g.p_i$ is determined
by a simple intraprocedural static analysis over instruction body (behavior) $g.s$: if $g.p_i$ is only read (and never written
or reduced), it is classified as read-only; if it is written but never read or reduced, it is write-only; otherwise it is read--write.
The atomic case is selected exactly when
$T_\mathrm{atom}(g.p_i)\neq\emptyset$.

\vspace{-0.3em}
\subsubsection{Program Conversion}
\label{sec:program-conversion}

The \ir{} 's program $\texttt{proc}~~ y^*, (x:\tau_x)^*~~;~~\texttt{do}~s$ is translated into an AMIR program $\texttt{proc}~~ y^*~~;~~\texttt{do}~s^\#$ where only control arguments are retained.
Non-control arguments $(x:\tau_x)^*$ are ignored during the conversion.
Control variables are preserved, since \ir{} and AMIR share the same control variables set $\mathbb{Y}$.
$T_s$ is applied to the program body to convert $s$ to $s^\#$.



\vspace{-0.3em}
\subsection{Abstract Machine Semantics}
\label{sec:amir-semantics}




\begin{definition}[Control–Value Environment]
\label{def:control-value-env}
Let $\mathbb{Z}$ be the set of control value and $\mathbb{Y}$ the set of control variables.
$\sigma \in \Sigma \;\triangleq\; \mathbb{Y}\to\mathbb{Z}$.
\end{definition}

\begin{definition}[Global Threads Collective]
\label{def:global-threads-collective}
Let $\mathbb{I}$ be the set of task IDs and $\mathbb{T}$ the set of all (local) thread IDs (thread collective).
The set of global thread IDs is $\mathbb{G}\triangleq \mathbb{I}\times\mathbb{T}$.
For each statement $s^\#$, thread mapping $f_{s^\#}:\Sigma \to \mathbb{T}$ returns the set of local threads that execute $s^\#$ under~$\sigma$ ($f_{s^\#}$ defined in Section~\ref{sec:collective-analysis}).
The scope mapping $\mathsf{scope}:\mathrm{Stmt}^\# \to \{ \text{CPU}, \text{CUDA} \}$ is initialized during the statement conversion to indicate whether each statement belongs to the CPU or CUDA scope.
Given $\sigma$ and $\imath\in\mathbb{I}$, define the corresponding global threads
\vspace{-0.2em}
\[
  g_{s^\#}(\sigma,\imath) \;\triangleq\; 
  \begin{cases}
        \{\,(\imath,t)\mid t\in f_{s^\#}(\sigma)\,\}\  & \text{if} \; \mathsf{scope}(s) = \text{CUDA} \\
        \mathbb{G} & \text{if} \; \mathsf{scope}(s) = \text{CPU}
  \end{cases}
\vspace{-0.1em}
\]
\end{definition}

\begin{definition}[Synchronization Environment]
\label{def:sync-env}
Let $\mathbb{B}$ be the set of barrier names and $\mathbb{X}$ the set of data variables.
Define
\vspace{-0.2em}
\[
  \mathbb{P}:\mathbb{B}\to\mathbb{Z}^n\to\mathcal{P}(\mathbb{N})
  \quad\text{and}\quad
  \mathbb{S}:\mathbb{G}\to\mathrm{QualTL}\to\mathrm{VL},
\vspace{-0.1em}
\]
where $\mathbb{P}(b, c^*)$ is the set of observed arrive counts at barrier access $b[c^*]$, and $\mathbb{S}(g, \tau_q)$ is the visibility level for the thread with global ID $g \in \mathbb{G}$ on a qualitative timeline $\tau_q$ (Figure~\ref{fig:am-syntax-def}).
Let
\[
  \mathbb{R}\ \triangleq\ \mathrm{QualTL}\times \mathbb{S} \times \mathbb{P}
\]
be the set of \emph{visibility records}, written (q, $\mathcal{s}$, p).
The synchronization environment is the map
\vspace{-0.1em}
\[
  \rho:\;(\mathbb{X}\cup\mathbb{B})\to\mathbb{Z}^n\ \to\ \mathcal{P}(\mathbb{R})\times\mathcal{P}(\mathbb{R})\times\mathbb{N}^2,
\vspace{-0.1em}
\]
sending a name and index to a tuple of (read visibility record set $r: \mathcal{P}(\mathbb{R})$, mutate visibility record set $m: \mathcal{P}(\mathbb{R})$, barrier count $b: \mathbb{N}^2$).
\end{definition}

The semantics follows a standard big-step style where judgments take the form \(\langle e^\#, \sigma \rangle \Downarrow v\) for expressions and \(\langle s^\#, \imath, \sigma, \rho \rangle \Downarrow \imath', \sigma', \rho'\) for statements, indicating that expression \(e^\#\) evaluates to value \(v\) in state \(\sigma\), and statement \(s^\#\) transforms state \(\sigma\) to \(\sigma'\), $\rho$ to $\rho'$, and $\imath$ to $\imath'$, respectively.

\input{tex_gpu/semantics}

\subsubsection{Expression semantics}
Figure~\ref{fig:big-step-semantics-exprs} gives big-step judgements \( \langle e,\sigma \rangle \Downarrow_{\tau} v \).
Integer and boolean literals evaluate to themselves (\textsc{C-Int}, \textsc{B-Const}); variables read from the store (\textsc{C-Read}).
Arithmetic evaluates both operands to integers and then applies \( \mathrm{op}\in\{+,-,*,/\} \) (\textsc{C-Aff}).
Boolean connectives apply the usual truth functions for \texttt{and}/\texttt{or} (\textsc{B-Log}), and comparisons \( \mathrm{op}\in\{==,<,\leq\} \) compare integer results to yield booleans (\textsc{B-Rel}).
Windows denote index sets \( \mathcal{I} \): a point yields \( \{i\} \) (\textsc{W-Point}); a range \( c_1..c_2 \) yields the inclusive set \( \{\, i\in\mathbb{Z}\mid i_1\le i \le i_2 \,\} \) (\textsc{W-Range}); and an \(n\)-tuple of windows denotes the Cartesian product \( W_1\times\cdots\times W_n \) (\textsc{W-Tuple}).
The only nonstandard part is \textsc{E\#-Acc}: given \( a\in\mathbb{X}\cup\mathbb{B} \) and a window tuple \( \mathbf{W} \), it \emph{unpacks} the window into the set of concrete accesses \( \{\, \{a\mapsto \vec{i}\} \mid \vec{i}\in \mathbf{W} \,\} \) of type $\mathbb{X} \cup \mathbb{B} \to \mathbb{Z}^*$.

\vspace{-0.4em}
\subsubsection{Statement semantics}
Figure~\ref{fig:big-step-semantics-loops} gives big-step operational semantics for a loop, sequencing, an explicit task–ID increment, and a guard. For $\texttt{for }y\text{ in }\texttt{loop}(c_{\mathrm{lo}},c_{\mathrm{hi}})\text{ do }s^\#$, the bounds evaluate in $\sigma$ to integers $m$ and $n$; the loop ranges over the \emph{half-open} interval $[m,n)$: when $m<n$ (\textsc{LoopStep}) the body executes once with $y\mapsto m$, producing $(\imath',\sigma',\rho')$, and the loop recurs on $(m{+}1)\dots n$; when $m\ge n$ (\textsc{LoopDone}) it terminates with no effect. Sequencing (\textsc{Seq}) threads the triple $(\imath,\sigma,\rho)$ from the first statement into the second.
The guard (\textsc{IF-True}/\textsc{IF-False}) evaluates $b$ in $\sigma$ and either executes $s^\#$ or behaves as no-op.
The only non-standard rule, \textsc{InctaskID}, increments the task identifier ($\imath'=\imath+1$) while leaving $\sigma$ and $\rho$ unchanged.

Figure~\ref{fig:big-step-semantics} shows the rest of statement semantics.
\textsf{Alloc} evaluates each extent $c_i$ to an integer $n_i$ (under $\Downarrow_{\mathbb{Z}}$), forms the index set $I=\{(v_1,\dots,v_k)\mid 0\le v_i<n_i\}$, and initializes the per-array-element synchronization environment state for $x$ by updating $\rho$ so that, for every $\mathbf{n}\in I$, $\rho(x,\mathbf{n})=(\emptyset,\emptyset,(0,0))$ (no visibility records, 0 arrives/awaits); the thread id $\imath$ and store $\sigma$ are unchanged. \textsf{ClearReads} first resolves the accessed locations via $\langle e^\#,\sigma\rangle\Downarrow_{\mathsf{Acc}}W_{e^\#}$ and, for each $(x,\mathbf{n})\in W_{e^\#}$, replaces $\rho(x,\mathbf{n})=(r,m,b) \in \mathbb{R}$ with $(\emptyset,m,b)$. \textsf{ClearMutates} is analogous but replaces it with $(r,\emptyset,b)$.
\textsc{RecordRead} and \textsc{RecordMutate} first evaluate the accessor \(e^{\#}\) and \(e^{\#}_{z}\) to access sets \(W_{e^{\#}}\) and \(W_{e^{\#}_{z}}\) (via \(\Downarrow_{\mathsf{Acc}}\)). They then update the visibility map \(\rho\) at each key \((x,\mathbf{n}) \in W_{e^{\#}}\) using the new record constructor $\mathcal{R}$ defined in Appendix~\ref{appendix:am-figures}.

\texttt{Fence}, \texttt{Arrive}, and \texttt{Await} update $\rho$ solely via $\mathrm{lift}(\rho,\lambda)$ (defined in Appendix~\ref{appendix:am-figures}), which applies $\lambda$ pointwise to both the read and mutate visibility record sets.
In the following, $\mathcal{W}$ decides when augmentation is permitted, $\mathcal{A}$ raises visibility to \texttt{fully\_ordered} or \texttt{temporally\_ordered}.
For \texttt{Fence}$(t,\tau_q^{\mathrm{pre}*},\tau_q^{\mathrm{full}*},\tau_q^{\mathrm{temp}*})$, $\lambda(r)=\mathcal{A}\!\left(r,\,g^{\mathrm{exec}*},\,\tau_q^{\mathrm{full}*},\,\tau_q^{\mathrm{temp}*}\right)$ iff the witness $\mathcal{W}\!\left(t,\,g^{\mathrm{exec}*},\,\tau_q^{\mathrm{pre}*},\,r\right)$ holds, otherwise $\lambda(r)=r$; thus visibility is raised by joining $f\vee f_{\mathrm{aug}}$. 
\texttt{Arrive}$(t,\tau_q^{\mathrm{pre}*},e_z^{\#*})$ and \texttt{Await}$(e_z^{\#},\tau_q^{\mathrm{full}*},\tau_q^{\mathrm{temp}*},n)$ are defined similarly in Appendix~\ref{appendix:am-figures}, updating the access visibility appropriately.

\subsubsection{Checks}

Figure~\ref{fig:big-step-semantics-checks} gives three pure checks that either signal $\text{error}$ or leave $(\imath,\sigma,\rho)$ unchanged. \textsc{CheckReads} evaluates $e^\#$ to accesses $W_{e^\#}$ and raises $\text{error}$ when some $(\_,\mathcal{s},\_)\in\rho(W_{e^\#}).r$ fails to supply visibility at least $\tau_v$ for all timelines $\tau_q\!\in\!\tau_q^*$. \textsc{CheckMutates} is analogous but ranges over mutate visibility record set $\rho(W_{e^\#}).m$. \textsc{CheckBarrier} evaluates $e_z^\#$ and signals $\text{error}$ iff some $(a,w)\in\rho(W_{e_z^\#}).b$ has $a\neq w$, i.e., the recorded \emph{arrive} and \emph{await} counts disagree.

\subsubsection{Program semantics}

Finally, abstract machine executes the program $\texttt{proc}~~ y^*~~;~~\texttt{do}~s^\#$.
The synchronization check is exposed to the user, who invokes the abstract machine interpreter with concrete integer values corresponding to the control variables $y^*$.
The control-value environment $\sigma : \mathbb{Y} \to \mathbb{Z}$ is initialized based on the user's input.
The statement $s^\#$ is then evaluated according to the semantics defined previously.
Note that the task id $\imath$ is initialized to $0$, and the synchronization $\rho$ is initially empty.

%% file: tex_gpu/grammar_abs_machine.tex
\begin{figure}[t]
\footnotesize
\centering
\arraycolsep=1.8pt\def\arraystretch{1.0}
\setlength{\tabcolsep}{2pt}
\begin{tabular}{rrll}
\toprule
  $\tau_v : \mathrm{VL}$ & $\Coloneqq$ &
  $\begin{array}{l}
    \texttt{fully\_ordered} $\,|\, $ \texttt{temporally\_ordered} \\ $\,|\, $
    \texttt{unordered} $\,|\, $ \texttt{atomic\_only} $\,|\, $ \texttt{invisible}
  \end{array}$
  & visibility levels \\
\midrule
  $e^\# : \mathrm{Expr}^\#$ & $\Coloneqq$ &
    $e_d^\# \, | \, e_z^\# \qquad$ where $e_d^\# \triangleq x[w^*]  \quad e_z^\# \triangleq z[w^*]$
    & barrier and data accesses \\
\midrule
  $s^\# : \Stmt^\#$ & $\Coloneqq$ &
        $s^\#_1 \texttt{;} s^\#_2$ & sequencing \\
 &$|$& \texttt{for $y$ in \texttt{loop}($c_{\mathrm{lo}}, c_{\mathrm{hi}}$) do $s^\#$} & loop \\
  &$|$& \texttt{if $b$ then $s^\#$} & guard \\
 &$|$& \texttt{IncTaskID} & increments the task ID \\
  &$|$& \texttt{Alloc($e^\#$)} & updates SyncEnv for Alloc \\
  &$|$& \texttt{RecordRead($\tau_v, t, e^\#, \tau_q, \tau_q^*, e_\mathrm{z}^{\#}$)} & updates SyncEnv with Read \\
  &$|$& \texttt{RecordMutate($\tau_v, t, e^\#, \tau_q, \tau_q^*, e_\mathrm{z}^{\#}$)} & updates SyncEnv with Mutate \\
  &$|$& \texttt{ClearReads($e^\#$)} & removes state added by \texttt{RecordRead} \\
  &$|$& \texttt{ClearMutates($e^\#$)} & removes state added by \texttt{RecordMutate} \\
  &$|$& \texttt{Fence(}$t, \tau_q^*, \tau_q^*, \tau_q^*$\texttt{)} & non-split barriers \\
  &$|$& \texttt{Arrive($t, \tau_q^*, e_\mathrm{z}^{\#*}$)} & arrive \\
  &$|$& \texttt{Await(}$e_\mathrm{z}^\#, \tau_q^*, \tau_q^*$, $n$\texttt{)} & await \\
  &$|$& \texttt{CheckReads($\tau_v, t, e^\#, \tau_q^*$)} & checks visibility of Reads in SyncEnv \\
  &$|$& \texttt{CheckMutates($\tau_v, t, e^\#, \tau_q^*$)} & checks visibility of Mutates in SyncEnv \\
  &$|$& \texttt{CheckBarrier($e_\mathrm{z}^\#$)} & checks the consistency of arrive and await \\
\midrule
$p^\# : \Proc^\#$ & $\Coloneqq$ &
  $\begin{array}{l}
    \texttt{proc}~~ y^* \\
    ~~\texttt{do}~s^\#
  \end{array}$ & arguments are control-only \\
  \bottomrule
\end{tabular}
\vspace{-0.5em}
\caption{Syntax of Abstract Machine IR (AMIR).
Variables, types, and expressions ($c,b,w$) use the same definition as \ir{}.
Visibility levels are denoted by the \emph{type} $\tau_v \in \mathrm{VL}$ and range over the five constructors listed above.
}
\vspace{-1.5em}
\label{fig:am-syntax-def}
\end{figure}


%% file: tex_gpu/convresion.tex
\newcommand{\Qcs}{\mathrm{Qcs}}
\newcommand{\Tcs}{\mathrm{Tcs}}
\newcommand{\cpuqual}{\texttt{cpu\_in\_order\_qual}}
\newcommand{\FenceCSpre}{\texttt{Fence(}true$,$ $\Qcs \cup \{\cpuqual\},$ $\Qcs,$ $\Tcs$\texttt{)}}
\newcommand{\FenceCSpost}{\texttt{Fence(}true$,$ $\Qcs,$ $\Qcs,$ $\Tcs$\texttt{)}}
\newcommand{\bigseq}{\mathop{\vcenter{\hbox{\Large$;$}}}\displaylimits}

\begin{figure}[t]
\centering
\footnotesize
\setlength{\tabcolsep}{4pt}
\begin{tabular}{ m{5.1cm} m{8.6cm} }
\toprule
\textbf{\ir{}} & \textbf{Abstract Machine IR} \\
\midrule

$s_1 ; s_2$ &
$T_s(s_1) ;~ T_s(s_2)$ \\ \midrule

\texttt{for $y$ in seq($c_{\mathrm{lo}},c_{\mathrm{hi}}$) do $s$} &
\texttt{for $y$ in loop($c_{\mathrm{lo}},c_{\mathrm{hi}}$) do $T_s(s)$} \\ \midrule

\texttt{for $y$ in cuda\_tasks($c_{\mathrm{lo}},c_{\mathrm{hi}}$) do $s$} &
\texttt{for $y$ in loop($c_{\mathrm{lo}},c_{\mathrm{hi}}$) do (IncTaskID; $T_s(s)$)} \\ \midrule

\texttt{for $y$ in cuda\_threads($c_{\mathrm{lo}},c_{\mathrm{hi}}, \tau_u$) do $s$} &
\texttt{for $y$ in loop($c_{\mathrm{lo}},c_{\mathrm{hi}}$) do $T_s(s)$} \\ \midrule

\textbf{$x[c^*]~\mathrel{\diamond}~ e$} \;\; \textit{where } $\diamond \in \{=,~\mathrel{+}=\}$ &
\begin{tabular}[t]{@{}l@{}}
$\bigseq_{e^\#_1 \in T_\mathrm{ex}(e)} \; \bigl(
\texttt{CheckMutates(}\tau_v^{\mathrm{pre}}, \texttt{false}, e^\#_1, \{T_{\mathrm{qin}}(e_1^\#) \}\texttt{)})\texttt{;}$\\
\hspace{1.31cm}$\texttt{RecordRead(}\tau_v^{\mathrm{post}}, \texttt{false}, e^\#_1, T_{\mathrm{qin}}(e_1^\#) , \emptyset, \emptyset \texttt{)} \bigr)$\\[3pt]
$\bigseq_{e^\#_2 \in T_\mathrm{ex}(x[c^*])} \; \bigl(
\texttt{CheckReads(}\tau_v^{\mathrm{pre}}, \texttt{false}, e^\#_2, \{T_{\mathrm{qin}}(e_2^\#) \}\texttt{)}) \texttt{;}$\\
\hspace{1.74cm}$\texttt{CheckMutates(}\tau_v^{\mathrm{pre}}, \texttt{false}, e^\#_2, \{T_{\mathrm{qin}}(e_2^\#) \}\texttt{)}) \texttt{;}$ \\
\hspace{1.74cm}$\texttt{ClearReads(}e^\#_2 \texttt{);}$ \\
\hspace{1.74cm}$\texttt{ClearMutates(}e^\#_2 \texttt{);}$ \\
\hspace{1.74cm}$\texttt{RecordMutate(}\tau_v^{\mathrm{post}}, \texttt{false}, e^\#_2, T_{\mathrm{qin}}(e_2^\#) , \emptyset, \emptyset \texttt{)}) \texttt{;} \bigr)$
\\
\text{, where } $\tau_v^{\mathrm{post}} = \texttt{fully-ordered}$ \text{ and }\\[2pt]
$\tau_v^{\mathrm{pre}} = \texttt{fully-ordered}$ if $\mathrel{+}=$, else $\texttt{temporally-ordered}$ \\
\end{tabular}
\\ \midrule

\texttt{if $b$ then $s$} &
\texttt{if $b$ then $T_s(s)$} \\ \midrule

\texttt{with CudaDeviceFunction(}$n,~e_\mathrm{w}^*$\texttt{) do $s$} &
\begin{tabular}[t]{@{}l@{}}
\FenceCSpre\texttt{;} \\
$T_s(s)$\texttt{;}\\
\FenceCSpost\texttt{;}\\[2pt]
, where $\mathrm{Qcs} \triangleq \mathrm{T_{qfull}}(\texttt{cuda\_stream\_sync})$, $\mathrm{Tcs} \triangleq \mathrm{T_{qtmp}}(\texttt{cuda\_stream\_sync})$
\end{tabular}
\\ \midrule

\texttt{with CudaWarps($n$, $n$, $r$) do $s$} &
\texttt{if $\mathrm{true}$ then $T_s(s)$} 
\\ \midrule

$g(c^*, e^*)~\texttt{>>}~e_\mathrm{z}$ & 
\begin{tabular}[t]{@{}l@{}}
$\bigseq_{e^\# \in T_\mathrm{ex}(e_z)}$\;
$\texttt{RecordRead(}\texttt{fully-ordered}, \texttt{false}, e^\#, \emptyset, \emptyset, T_e(e_\mathrm{z})\texttt{)}$ \\[2pt]
$\bigseq_{i=0}^{\mathrm{len}(e^*)} T_\mathrm{arg}(g.p_i, \, e_i, \, e_\mathrm{z})$\\
\end{tabular}
\\ \midrule

\texttt{$x:\tau_x[c^*] ~@~ \pi_\mathrm{d}$} &
\texttt{Alloc(}$x[c^*]$\texttt{)} \\ \midrule

\texttt{$z:\tau_z[c^*] ~@~ \pi_\mathrm{z}$} &
\texttt{Alloc(}$z[c^*]$\texttt{)}
\\ \midrule

\texttt{free $x$} & 
\begin{tabular}[t]{@{}l@{}}
$\bigseq_{e^\# \in T_\mathrm{smem}(x[:^*])} \; \bigl(
\texttt{CheckReads(}\tau_v^{\mathrm{free}}, \texttt{false}, e^\#, \{T_{\mathrm{qin}}(e^\#)\}\texttt{)}) \texttt{;}$\\
\hspace{1.99cm}$\texttt{CheckMutates(}\tau_v^{\mathrm{free}}, \texttt{false}, e^\#, \{T_{\mathrm{qin}}(e^\#)\}\texttt{)}) \texttt{;}\bigr)$ \\[2pt]
\end{tabular}
\\ \midrule
\texttt{free $z$} &  
\begin{tabular}[t]{@{}l@{}}
\texttt{CheckBarriers(} $T_e(z[:^*])$ \texttt{);}\\
$\bigseq_{e^\# \in T_\mathrm{smem}(z[:^*])} \; \bigl(
\texttt{CheckReads(}\tau_v^{\mathrm{free}}, \texttt{false}, e^\#, \{T_{\mathrm{qin}}(e^\#)\}\texttt{)}) \texttt{;}$\\
\hspace{1.95cm}$\texttt{CheckMutates(}\tau_v^{\mathrm{free}}, \texttt{false}, e^\#, \{T_{\mathrm{qin}}(e^\#) \}\texttt{)}) \texttt{;}\bigr)$ \\[2pt]
\end{tabular}
\\ \midrule

\texttt{Fence(}$\tau_s^{\mathrm{pre}}, \tau_s^{\mathrm{post}}$\texttt{)} &
\texttt{Fence(}$\mathrm{T_{tr}}(\tau_s^{\mathrm{pre}}),~
\mathrm{T_{qfull}}(\tau_s^{\mathrm{pre}}),~
\mathrm{T_{qfull}}(\tau_s^{\mathrm{post}}),~
\mathrm{T_{qtmp}}(\tau_s^{\mathrm{post}})$\texttt{)}
\\ \midrule

\texttt{Arrive(}$\tau_s, n$\texttt{)} \texttt{>>} $e_\mathrm{z}^*$ &
\begin{tabular}[t]{@{}l@{}}
$\bigseq_{e^\# \in T_\mathrm{ex}(e_\mathrm{z}^*)}\;
\texttt{RecordRead(}\texttt{fully-ordered}, \texttt{false}, e^\#, \mathrm{T_{qarr}}(\tau_s), \emptyset, \emptyset$\texttt{)}\\[2pt]
\texttt{Arrive(}$\mathrm{T_{tr}}(\tau_s),~
\mathrm{T_{qfull}}(\tau_s),~
T_e(e_\mathrm{z}^*)$\texttt{);} \\
\end{tabular}
\\ \midrule

\texttt{Await(}$e_\mathrm{z}, \tau_s, n$\texttt{)} &
\begin{tabular}[t]{@{}l@{}}
$\bigseq_{e^\# \in T_\mathrm{ex}(e_z)}$\;
$\texttt{RecordRead(}\texttt{fully-ordered}, \texttt{false}, e^\#, T_{\mathrm{qin}}(e^\#), \emptyset, \emptyset\texttt{)}$ \\[2pt]
\texttt{Await(}$\mathrm{T_e}(e_\mathrm{z}),~
\mathrm{T_{qfull}}(\tau_s),~
\mathrm{T_{qtmp}}(\tau_s),~n$\texttt{);}
\\
\end{tabular}
\\
\bottomrule
\end{tabular}
\vspace{-0.5em}
\caption{$T_s : \Stmt \to \Stmt^\#$ (statement conversion). All \emph{conversion} functions use the $T\_\cdot$ prefix (Figures~\ref{fig:statement_conversion_helpers}-\ref{fig:targ_conversion_helpers}).
$\tau_v^{\mathrm{free}} = \texttt{temporally-ordered}$
}
\label{fig:statement_conversion}
\end{figure}


\newcommand{\SyncTL}{\mathrm{SyncTL}}
\newcommand{\QualTL}{\mathrm{QualTL}}
\newcommand{\ZExpr}{\mathrm{ZExpr}}

\newcommand{\Targ}{T_\mathrm{arg}}
\newcommand{\VLfull}{\texttt{full\_ordered}}
\newcommand{\VLtmp}{\texttt{temporal\_ordered}}
\newcommand{\VLun}{\texttt{unordered}}
\newcommand{\VLatom}{\texttt{atomic\_only}}
\newcommand{\vpost}{\tau_v^\mathrm{post}}

%% file: tex_gpu/semantics.tex



\begin{figure}[t]
\centering
\small
\begin{minipage}[h]{\linewidth}
\centering
\begin{mathpar}
\inferrule*[right=C-Int]{ }{
  \langle n,\sigma\rangle \Downarrow_{\mathbb{Z}} n }

\inferrule*[right=C-Read]{ \sigma(y)=n }{
  \langle y,\sigma\rangle \Downarrow_{\mathbb{Z}} n }

\inferrule*[right=C-Aff]{ 
  \langle c_1,\sigma\rangle \Downarrow_{\mathbb{Z}} n_1 \\
  \langle c_2,\sigma\rangle \Downarrow_{\mathbb{Z}} n_2 \\
  \,\mathrm{op} \in \{ {+},{-},*,/\}
}{
  \langle c_1 \,\mathrm{op}\, c_2,\sigma\rangle \Downarrow_{\mathbb{Z}} (n_1 \,\mathrm{op}\, n_2)
}

\inferrule*[right=B-Const]{ }{
  \langle t,\sigma\rangle \Downarrow_{\mathsf{Bool}} t
}

\inferrule*[right=B-Log]{
  \langle b_1,\sigma\rangle \Downarrow_{\mathsf{Bool}} t_1 \\
  \langle b_2,\sigma\rangle \Downarrow_{\mathsf{Bool}} t_2 \\
  \,\mathrm{op} \in \{\texttt{and},\texttt{or}\}
}{
  \langle b_1 \,\mathrm{op}\, b_2,\sigma\rangle \Downarrow_{\mathsf{Bool}} \widehat{\mathrm{op}}(t_1,t_2)
}

\inferrule*[right=B-Rel]{
  \langle c_1,\sigma\rangle \Downarrow_{\mathbb{Z}} n_1 \\
  \langle c_2,\sigma\rangle \Downarrow_{\mathbb{Z}} n_2 \\
  \,\mathrm{op} \in \{ {==},{<},{\leq} \}
}{
  \langle c_1 \,\mathrm{op}\, c_2,\sigma\rangle \Downarrow_{\mathsf{Bool}} (n_1 \,\mathrm{op}\, n_2)
}

\inferrule*[right=W-Point]{
  \langle c,\sigma\rangle \Downarrow_{\mathbb{Z}} i
}{
  \langle c,\sigma\rangle \Downarrow_{\mathcal{I}} \{\,i\,\}
}

\inferrule*[right=W-Range]{
  \langle c_1,\sigma\rangle \Downarrow_{\mathbb{Z}} i_1 \\
  \langle c_2,\sigma\rangle \Downarrow_{\mathbb{Z}} i_2
}{
  \langle c_1..c_2,\sigma\rangle \Downarrow_{\mathcal{I}}
  \{\, i \in \mathbb{Z} \mid i_1 \le i \le i_2 \,\}
}

\inferrule*[right=W-Tuple]{
  \forall j\in\{1..n\}.\; \langle w_j,\sigma\rangle \Downarrow_{\mathcal{I}} W_j
}{
  \langle w_1,\ldots,w_n,\sigma\rangle \Downarrow_{\mathcal{I}^n} W_1 \times \cdots \times W_n
}

\inferrule*[right=E\#-Acc]{
  a \in \mathbb{X} \cup \mathbb{B} \\
  \langle w_1,\ldots,w_n,\sigma\rangle \Downarrow_{\mathcal{I}^n} \mathbf{W}
}{
  \langle a[w_1,\ldots,w_n],\sigma\rangle \Downarrow_{\mathsf{Acc}}
  \{\, \{a \mapsto \vec{i}\} \mid \vec{i} \in \mathbf{W} \,\}
}

\end{mathpar}
\end{minipage}
\vspace{-0.5em}
\caption{The big-step operational semantics for expressions.}
\vspace{-0.7em}
\label{fig:big-step-semantics-exprs}
\end{figure}

\begin{figure}[t]
\centering
\small
\begin{minipage}[h]{\linewidth}
\centering
\begin{mathpar}

\inferrule*[right=CheckReads]{
  \exists (\_, \mathcal{s}, \_) \in \rho(W_{e^\#}).r \,
  \bigr((t = \text{false} \wedge
  \exists g \in g_{s^\#}(\sigma, \imath).\, \forall \tau_q \in \tau_q^* \mid \mathcal{s}(g, \tau_q) \nsucceq \tau_v) \\
  \langle e^\#, \sigma \rangle \Downarrow_{\mathsf{Acc}} W_{e^\#} \hspace{1.75cm}  
  \vee (t = \text{true} \wedge 
  \forall g \in g_{s^\#}(\sigma, \imath).\, \forall \tau_q \in \tau_q^* \mid \mathcal{s}(g, \tau_q) \nsucceq \tau_v)\bigl)
  \Downarrow_{\mathsf{Bool}} \text{true}
}{
  \langle \text{CheckReads(} \tau_v, t, e^\#, \tau_q^* \text{)}, \imath, \sigma, \rho \rangle
  \Downarrow \text{error}
}

\inferrule*[right=CheckMutates]{
  \exists (\_, \mathcal{s}, \_) \in \rho(W_{e^\#}).m \,
  \bigr((t = \text{false} \wedge
  \exists g \in g_{s^\#}(\sigma, \imath).\, \forall \tau_q \in \tau_q^* \mid \mathcal{s}(g, \tau_q) \nsucceq \tau_v) \\
  \langle e^\#, \sigma \rangle \Downarrow_{\mathsf{Acc}} W_{e^\#}
  \hspace{1.8cm} \vee
  (t = \text{true} \wedge 
  \forall g \in g_{s^\#}(\sigma, \imath).\, \forall \tau_q \in \tau_q^* \mid \mathcal{s}(g, \tau_q) \nsucceq \tau_v)\bigl)
  \Downarrow_{\mathsf{Bool}} \text{true}
}{
  \langle \text{CheckMutates(} \tau_v, t, e^\#, \tau_q^* \text{)}, \imath, \sigma, \rho \rangle
  \Downarrow \text{error}
}

\inferrule*[right=CheckBarrier]{
  \langle e_z^\#, \sigma \rangle \Downarrow_{\mathsf{Acc}} W_{e_z^\#} \\
  (\exists (a,w) \in \rho(W_{e_z^\#}).b \mid  a \neq w) \Downarrow_{\mathsf{Bool}} \text{true}
}{
  \langle \text{CheckBarrier(}e_z^\#\text{)},\, \imath,\, \sigma,\, \rho \rangle
  \Downarrow \text{error}
}

\end{mathpar}
\end{minipage}
\vspace{-0.7em}
\caption{Big-step rules for access visibility and barrier consistency. Successful checks leave $(\imath,\sigma,\rho)$ unchanged; any violation reduces the configuration to $\text{error}$ state.
}
\vspace{-1.3em}
\label{fig:big-step-semantics-checks}
\end{figure}

%% file: tex_gpu/04_impl_results.tex
\section{Performance Results}
\label{sec:results}

\begin{figure}[t]
\centering

\includegraphics[width=0.45\linewidth]{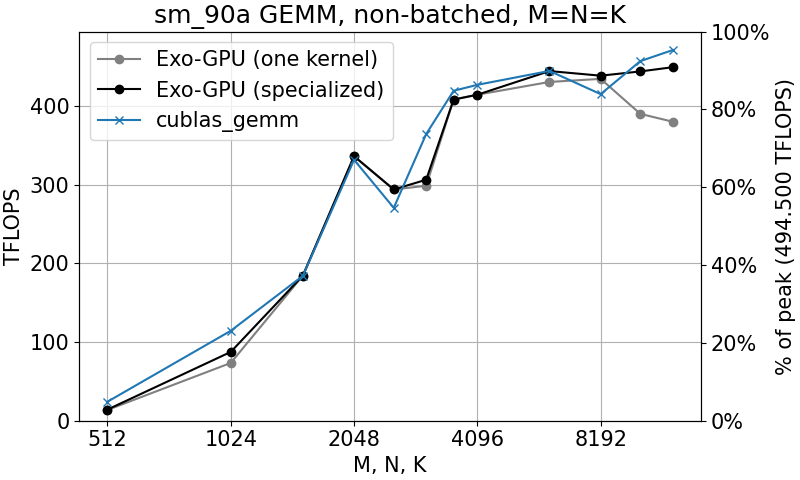}
\includegraphics[width=0.45\linewidth]{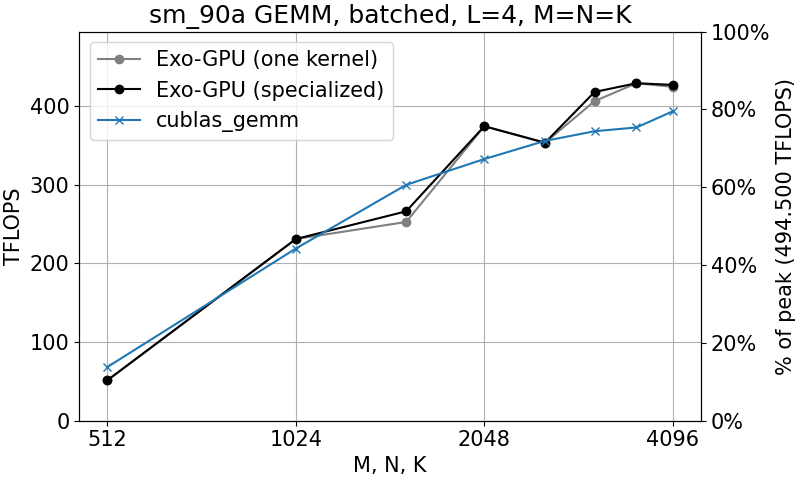}

\includegraphics[width=0.45\linewidth]{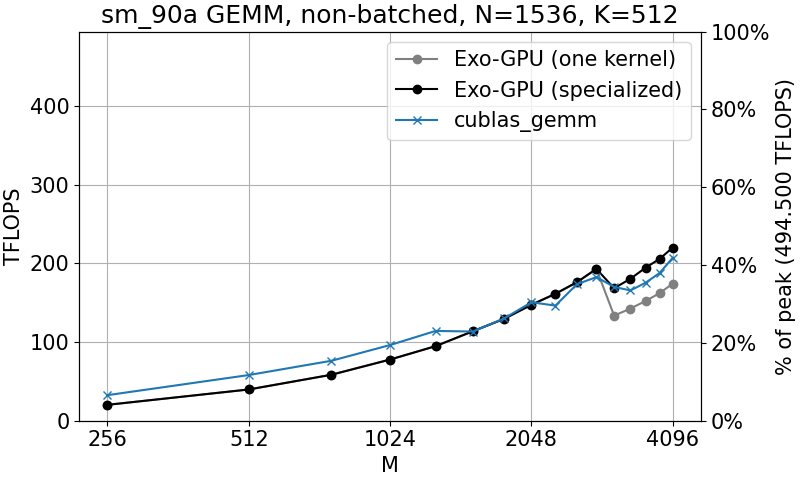}
\includegraphics[width=0.45\linewidth]{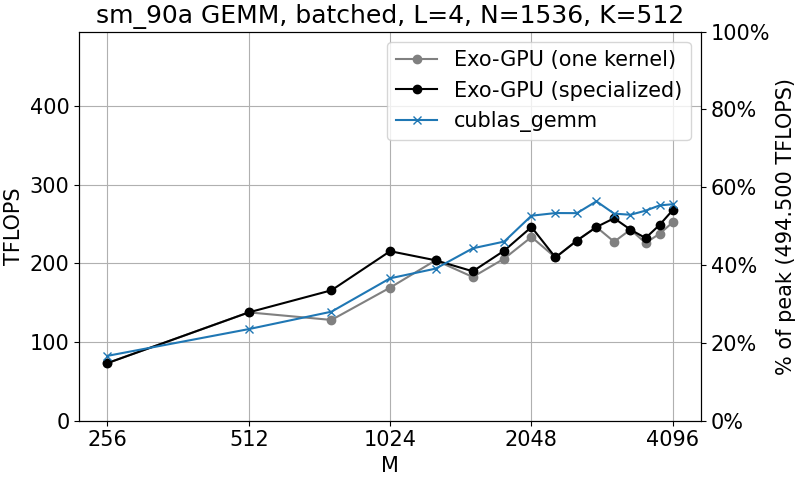}

\includegraphics[width=0.45\linewidth]{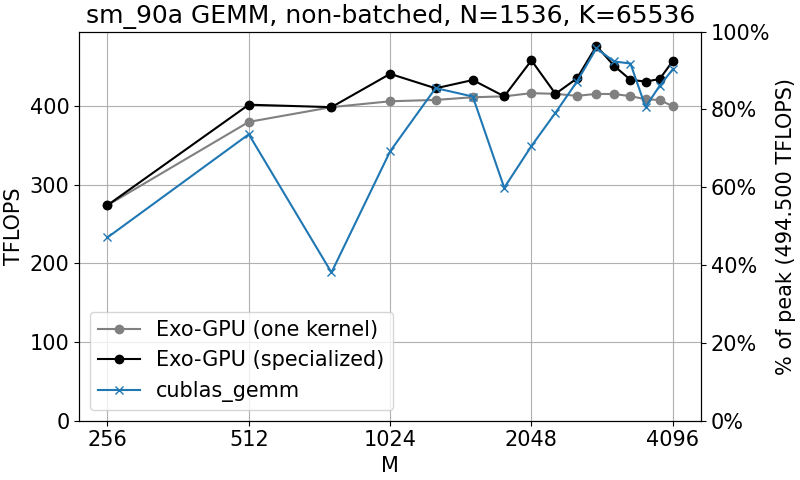}
\includegraphics[width=0.45\linewidth]{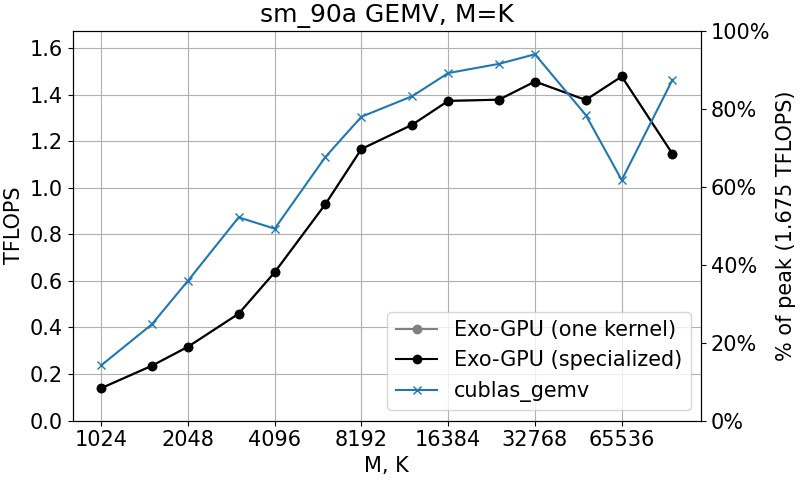}

\caption{Performance by varying problem size on a sm\_90a GPU.}
\label{fig:main_results}
\end{figure}



We implemented tf32-precision GEMM kernels for the sm\_90a architecture, using \texttt{wgmma} and TMA instructions.
We also implemented a full-precision fp32 GEMV kernel using warp shuffles and scalar math \cite{cuda_hgemv}; this did not use the accelerator instructions.
We compared the \name{} kernels to cuBLAS's \texttt{cublasSgemm}, \texttt{cublasSgemmStridedBatched}, and \texttt{cublasSgemv} functions, all with the math mode set to \texttt{CUBLAS\_TF32\_TENSOR\_OP\_MATH}.
%
Our sm\_90a-compatible
test hardware contained an NVIDIA H100 80GB SXM5 GPU and an AMD EPYC 9454 48-Core Processor, and we used nvcc \texttt{Build cuda\_12.3.r12.3/compiler.33492891\_0} to compile the CUDA C++ sources generated by \name{}.
For GEMM, a compute-bound workload, theoretical peak is 494.5 TFLOPS \cite{h100_gpu}.
For GEMV, a memory-bound workload, we derive the theoretical peak from the memory bandwidth: $3.35\frac{\mathrm{TB}}{s} \frac{2\;\mathrm{FLOP}}{4\;\mathrm{B}} = 1.675\frac{\mathrm{TFLOP}}{s}$, with one multiply and one add done for each 4 byte tf32 element loaded from the matrix (the bandwidth from loading the vector is amortized).


We tested each problem size and hardware combination over 105 iterations: 5 warm-up iterations, and 100 timed iterations, with the pcg3d hash algorithm \cite{pcg3d} used to generate ``random'' input matrices.
Each iteration tested, in a randomized order, the cuBLAS kernel and all applicable \name{} kernel variants for that problem size.
In particular, we tested non-split-k and split-k variants of the GEMM kernel, with the split-k kernels using \texttt{cp.reduce.async.bulk} (TMA) and split factors of 1, 2, 4, 8, and 16.
For each (problem size, hardware) combination, we collected 100 time samples per tested kernel and reported the mean of the inter-quartile range (middle 50 samples).
For each plot in Figure \ref{fig:main_results}, we report the best-performing \name{} kernel for each problem size (plotted as ``Exo-GPU (specialized)'') and the single best-performing \name{} kernel (``Exo-GPU (one kernel)'').

For large, square matrix GEMM problem sizes, we see very similar performance between \name{} and CUBLAS; this is expected, as \name{} is able to express all optimizations required for a peak-performance sm\_90a gemm: warp specialization, cluster \& TMA multicast support, \texttt{wgmma} support, and split-barrier synchronization using commit group, mbarrier, and cluster sync mechanisms.
For more unusual GEMM problem sizes, such as small matrices and \texttt{K=65536} workloads, we see more variation between \name{} and CUBLAS, possibly due to different problem size thresholds for observing wave quantization effects.
For GEMV, \name{} and CUBLAS follow a similar performance trend, with CUBLAS having an almost-consistent performance advantage.



\begin{wrapfigure}{r}{0.5\textwidth}
\centering
\includegraphics[width=\linewidth]{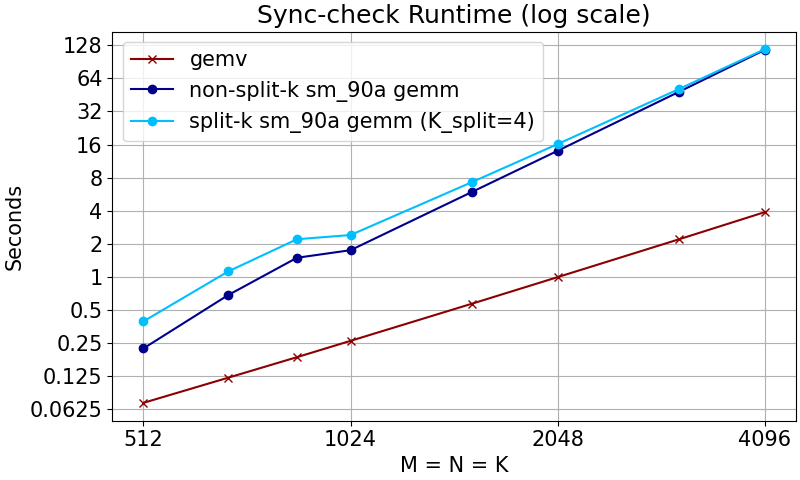}
\vspace{-2em}
\caption{Sync-check runtime by problem size.}
\vspace{-0.9em}
\label{fig:am-performance}
\end{wrapfigure}

\textbf{Abstract Machine performance.}
We report the performance of the synchronization checking alone, which is implemented as an interpreter for the abstract machine (Section~\ref{sec:abstract-machine}).
We benchmarked using only one thread of a laptop with an ``\texttt{11th Gen Intel(R) Core(TM) i7-11850H @ 2.50GHz}'' CPU.
In Figure~\ref{fig:am-performance}, we report times for concrete square matrix problem sizes for the GEMV kernel and the sm\_90a GEMM kernel described earlier (both the split-k and non-split-k versions).
The batch size \texttt{L} is 1.
We use one untimed warmup run for each problem size, and report the average of five timed runs, as reported by the \texttt{timeit} module of Python 3.10.12.

%

We observe that runtimes are linear in the number of memory operations interpreted.
The number of memory operations is cubic in the problem size for GEMM, and quadratic for GEMV.
We see this pattern followed cleanly by the measured trend (e.g. 2/16/128 seconds for GEMM for 1024/2048/4096 problem size), indicating no unexpected overhead for large problem sizes.
The bump for small GEMM problem sizes is due to imperfect division by the per-cluster tile size.
Even if the expected problem size for a deployed kernel is large, during development of an \name{} kernel, it generally suffices to test only with a small problem size (e.g. 768) for immediate feedback on bugs, with the option of testing the finalized kernel with the true, larger problem size for additional confidence.

%% file: tex_gpu/05_related_work.tex
\section{Related Work}

\subsection{GPU Kernel Authoring Languages}

While CUDA offers fine-grained control, it remains low-level and error-prone.
To address this, various approaches have emerged at different abstraction levels.

\vspace{-0.3em}
\paragraph{Higher-level abstractions:}
Tile-level DSLs like Triton \cite{triton} and Pallas \cite{pallas} abstract away low-level details by enabling programmers to express computations at the block level, with compilers automatically handling the thread-level mapping.
However, these abstractions can hide performance-critical details. When Triton performance fell far behind state-of-the-art for Blackwell GPUs, they had to invent Gluon, a lower-level language that gives users direct control over tile layouts, memory allocation, data movement, and asynchrony, essentially abandoning the higher-level automation to regain the fine-grained control necessary for peak performance.
Rather than starting with automation that may later prove inadequate, \name{} deliberately maintains a thin abstraction layer over CUDA.
Performance-critical constructs, including control flow and data copies across the memory hierarchy, are made explicit statements in \name{}'s imperative object code.

\vspace{-0.3em}
\paragraph{Template-based libraries:}
CUTLASS~\cite{cutlass} and ThunderKittens~\cite{thunderkittens} are C++ template-based libraries that raise the abstraction of GPU programming while preserving performance.
Instead of hiding CUDA behind rigid abstractions, they provide composable, architecture-aware primitives that integrate with raw CUDA code.
This approach lets programmers work at the tile level when convenient, yet seamlessly drop to thread-level operations when fine control is needed.
CuTe~\cite{cute}, introduced in CUTLASS 3.0, offers flexible abstraction for defining tensor layouts and transformations.
While these libraries are effective engineering toolkits, they do not provide safety guarantees or whole-program analyses for asynchronous instructions and synchronization statements.

Achieving both peak performance and safety remains a fundamental challenge.
Existing approaches either provide limited safety guarantees or lack support for modern GPU features.
For instance, Descend~\cite{descend} introduces Rust-like borrow checking to ensure race freedom but does not support key GPU features such as tensor cores and split barriers.
The aforementioned tile- and template-based languages are practical tools, yet do not offer safety guarantees.
This motivated the collective analysis (Section~\ref{sec:collective-analysis}) and synchronization checking via abstract machine (Section~\ref{sec:abstract-machine}).

\subsection{User-schedulable Languages}


User-schedulable languages (USLs) describe optimization as rewriting programs into new programs that compute the same result, but execute much faster. Rather than relying on opaque compiler heuristics, programmers directly control these rewrites by specifying scheduling operations.
Halide~\cite{Halide:SIGGRAPH,Halide:CACM} popularized the idea of USLs, and subsequent Halide-inspired languages explored different target applications and abstraction for a given domain~\cite{Legion,TVM,TensorComprehensions,TACO,GraphIt,SWIRL,FireIron,Taichi,exo,lift2017}.
For instance, TVM offers accelerator abstraction through \texttt{tensorize}, TACO separates tensor definitions from sparse formats, and Taichi abstracts memory layout and management with \texttt{fields}.


Several systems have extended user-scheduling to support modern GPU features. TVM's TensorIR~\cite{tensorir} provides intrinsics for manipulating low-level constructs like split barriers, while Cypress~\cite{cypress} lets programmers describe computations as tasks and map them to different resources.
\name{} differs from these approaches in two key ways. First, \name{} exposes low-level control to the programmer; there is little hidden control flow in \ir{}. In this sense, \name{} follows a WYSIWYG (what you see is what you get) philosophy: the structure of the program directly reflects its execution. This contrasts with Cypress, which hides warp specialization as an internal compiler optimization, and with TensorIR, which focuses on automatically generating schedules with lowering-based intrinsics.
Second, \name{} guarantees sequential-parallel equivalence through a synchronization checker that detects data races, hazards, and other concurrency bugs.
%





\subsection{Exo vs. \name{}}

\name{} fundamentally extends existing the Exo language by introducing parallel semantics.
While Exo~\cite{exo, exo2} is a high-performance USL that achieves peak performance on many vector machines and accelerators, it was designed to exploit only the \emph{implicit parallelism} exposed by hardware.
Exo's scheduling reasoning operates exclusively under a sequential execution model.
This approach suffices for vector machines and matrix accelerators because CPU frontends maintain instruction ordering; even with out-of-order execution and reorder buffers, the hardware guarantees that parallel execution produces results equivalent to sequential execution.
Therefore, the compiler needs only to verify program equivalence when interpreting object code sequentially.
Even within this sequential framework, Exo's optimizations could exploit implicit hardware parallelism through techniques like software pipelining~\cite{exo, exo2}.

In contrast, GPUs expose \emph{explicit data- and instruction-parallelism} to software.
Unlike traditional CPUs and accelerators that maintain the illusion of sequential execution through their frontend, modern GPUs shift this responsibility entirely to software.
For Exo, this shift requires reasoning about parallel execution at the object code level and establishing equivalence not just within programming models, but across sequential and parallel paradigms.
We address this fundamental challenge through two key contributions. First, we extend Exo's object code and scheduling language to expose user-controlled parallel loops and synchronization features.
Second, we provide safety guarantees by establishing sequential-parallel equivalence through the abstract machine.
This completes the equivalence chain from the simple, initial Exo program through to its optimized, parallel implementation in \name{}, ensuring correctness at every transformation step.

%% file: tex_gpu/06_limitations.tex
\section{Limitations and Future Work}
\label{sec:limitations}




\begin{wrapfigure}{r}{0.3\textwidth}
\vspace{-5mm}
\begin{minted}[fontsize=\footnotesize,breaklines]{python}
tmp : f32[8] @ tmp_mem
for i in seq(0, 8):
  if i < 6: # removing is safe
    tmp[i] = 0
for i in seq(0, 8):
  if i < 6:
    C[i] = tmp[i]
\end{minted}
\vspace{-5mm}
\end{wrapfigure}
To optimize Hopper GEMMs (Figure~\ref{fig:main_results}), we had to apply two rewrite steps that were out-of-scope of Exo's sequential safety checks.
These patterns arise when modeling TMA's asynchronous copy instruction, which supports default behavior for out-of-bounds reads and writes.
Inset right figure shows a minimal example that illustrates the problem.
Even though removing the first guard is safe since the \texttt{tmp[6]} and \texttt{tmp[7]} are not being read by \texttt{C} in the second loop thanks to the guard,
Exo's dependency-based checks treat these writes as semantically relevant and cannot prove  they are observationally redundant.
We marked these transformations as \emph{unsafe} in our schedules, despite being semantically sound.
To address this issue, we need to extend Exo's sequential scheduling rewrites with value-sensitivity.

Our ability to reason about parallelism is incomplete in several ways.
Our programming model and output CUDA synchronization code is generated based on our understanding of PTX as documented by NVIDIA in natural language \cite{ptx}.
We do not show that our generated code is correct with respect to any formal CUDA semantics.
We currently do not support Blackwell Tensor Cores (\texttt{tcgen05}), nor do we support platforms not providing CUDA C++ and inline PTX support.
Finally, our current synchronization checking approach -- translation to abstract machine IR followed by software interpretation -- supports only concrete problem sizes and limits compiler performance.
This approach may be improved with static analysis of the abstract machine program.

Nevertheless, we believe that our approach---exposing detailed control to programmers while verifying its safety---offers a promising path forward for GPU programming.
By treating parallelism and synchronization as analyzable annotations on sequential code, \name{} provides both the transparency and static guarantees that high-performance GPU kernels demand.

%% file: tex_gpu/acknowledgements.tex
We are thankful to William Brandon, Amanda Liu, Luca Musk, Jason Ng, Kevin Qian, and Philippe Tillet for their insightful discussions and early feedback.
This research was funded in part by the U.S. Government under the DARPA MOCHA program (HR001125CE038) and NSF award CCF-2217064, and by a gift from Jane Street.

%% file: tex_gpu/9_gpuir_types.tex
\section{\ir{} Type Definitions}
\label{sec:gpuir-type-figures}

\input{tex_gpu/grammar_timelines}

%% file: tex_gpu/grammar_timelines.tex
\begin{figure}[th]
\footnotesize
\centering
\arraycolsep=1.8pt\def\arraystretch{1.0}
\setlength{\tabcolsep}{2pt}
\begin{tabular}{rrll}
\toprule
$\tau_x : \mathrm{DataType}$ & $\Coloneqq$ &
  \texttt{f32} & 32-bit floating point \\
  &|& \texttt{f64} & 64-bit floating point \\
  &|& \texttt{f16} & 16-bit floating point \\
  &|& \texttt{i32} & 32-bit integer \\
  &|& \texttt{ui16} & 16-bit unsigned integer \\
  &|& \texttt{i8} & 8-bit integer \\
  &|& \texttt{ui8} & 8-bit unsigned integer \\
  [3pt]
$\tau_z : \mathrm{BarrierType}$ & $\Coloneqq$ &
  \texttt{barrier} & non-explicitly guarded barrier \\
  &|& \texttt{barrier(z)} & explicitly guarded barrier \\
\bottomrule
\end{tabular}
\caption{List of variable types}
\label{fig:variable-types}
\end{figure}

\begin{figure}[th]
\footnotesize
\centering
\arraycolsep=1.8pt\def\arraystretch{1.0}
\setlength{\tabcolsep}{2pt}
\begin{tabular}{rrlll}
\toprule
& & & \emph{domain} & \emph{box} \\
$\tau_u : \mathrm{CollUnit} $ & $\Coloneqq$ &
  \texttt{standalone\_thread} & \texttt{(1,)} & \texttt{(1,)} \\
  &|& \texttt{$n_1$ * cuda\_thread} & \texttt{(blockDim,)} & \texttt{($n_1$,)} \\
  &|& \texttt{cuda\_quadpair} & \texttt{(blockDim/16, 16)} & \texttt{(2, 4)} \\
  &|& \texttt{$n_1$ * cuda\_warp} & \texttt{(blockDim,)} & \texttt{($n_1$ * 32,)} \\
  &|& \texttt{$n_1$ * cuda\_warpgroup} & \texttt{(blockDim,)} & \texttt{($n_1$ * 128,)} \\
  &|& \texttt{$n_1$ * cuda\_threads\_strided($n_2$, $n_3$)} & \texttt{(blockDim/$n_3$, $n_3$)} & \texttt{($n_1$, $n_2$)} \\
  &|& \texttt{$n_1$ * cuda\_warp\_in\_cluster} & \texttt{(clusterDim, blockDim)} & \texttt{($n_1$, 32)} \\
  &|& \texttt{$n_1$ * cuda\_cta\_in\_cluster} & \texttt{(clusterDim * blockDim,)} & \texttt{($n_1$ * blockDim,)} \\
  &|& \texttt{cuda\_cluster} & \texttt{(clusterDim * blockDim,)} & \texttt{(clusterDim * blockDim,)} \\
  &|& \texttt{$n_1$ * cuda\_cta\_in\_cluster\_strided($n_3$)} & \texttt{(ClusterDim/$n_3$, $n_3$, blockDim)} & \texttt{($n_1$, 1, blockDim)} \\
  &|& \texttt{$n_1$ * cuda\_warp\_in\_cluster\_strided($n_3$)} & \texttt{(clusterDim/$n_3$, $n_3$, blockDim)} & \texttt{($n_1$, 1, 32)} \\
  &|& \texttt{cuda\_agnostic\_sub\_cta} & \texttt{(clusterDim, blockDim)} & \texttt{(1, \_)} \\
  &|& \texttt{cuda\_agnostic\_intact\_cta} & \texttt{(clusterDim, blockDim)} & \texttt{(\_, blockDim)} \\
\bottomrule
\end{tabular}
\caption{Collective units defined in the frontend language.
These are parameterized by a collective type $\delta$, which consists of the \emph{domain} and \emph{box} shown here.}
\label{fig:CollUnit}
\end{figure}

\begin{figure}[th]
\footnotesize
\centering
\arraycolsep=1.8pt\def\arraystretch{1.0}
\setlength{\tabcolsep}{2pt}
\begin{tabular}{rrll}
\toprule
$\pi_\mathrm{d} : \mathrm{DataMemory}$ & $\Coloneqq$ &
  \texttt{CudaBasicDeviceVisible} & base type for CUDA-visible allocations \\
  &|& \texttt{CudaBasicSmem} & base type for CUDA SMEM allocations \\
  &|& \texttt{CudaDeviceVisibleAtomicity16B} & base type for 16B-atomicity layout CUDA-visible allocations \\
  &|& \texttt{CudaDeviceVisibleLinear} & base type for linear-order CUDA-visible allocations \\
  &|& \texttt{CudaGridConstant} & CUDA grid constant memory \\
  &|& \texttt{CudaGmemLinear} & CUDA global memory, linear-order \\
  &|& \texttt{CudaSmemAtomicity16B} & any CUDA shared memory with 16B-atomicity layout \\
  &|& \texttt{CudaSmemLinear} & CUDA shared memory with linear-order layout \\
  &|& \texttt{CudaRmem} & per-thread CUDA registers \\
  &|& \texttt{Sm80\_RmemMatrixA(M, K)} & matrix tile for sm\_80+ warp MMA A operand \\
  &|& \texttt{Sm80\_RmemMatrixB(N, K)} & matrix tile for sm\_80+ warp MMA B operand \\
  &|& \texttt{Sm80\_RmemMatrixD(M, N)} & matrix tile for sm\_80+ MMA acumulator operands \\
  &|& \texttt{Sm90\_SmemSwizzled(B)} & B-byte swizzled shared memory, in TMA/wgmma-accepted format \\
  &|& \texttt{Sm90\_RmemMatrixA(M)} & register tile for wgmma A operand (not implemented) \\
  &|& \texttt{Sm90\_RmemMatrixD(M, N)} & register tile for wgmma D operand \\
\bottomrule
\end{tabular}
\caption{CUDA Memory Types -- Data Allocations (pre-existing Exo Memory not listed)}
\label{fig:CudaMemory}
\end{figure}

\begin{figure}[th]
\footnotesize
\centering
\arraycolsep=1.8pt\def\arraystretch{1.0}
\setlength{\tabcolsep}{2pt}
\begin{tabular}{rrll}
\toprule
$\pi_\mathrm{z} : \mathrm{BarrierComp}$ & $\Coloneqq$ &
  \texttt{CudaDeviceBarrier} & base type for CUDA-allocated barriers \\
  &|& \texttt{CudaMbarrier} & CUDA \texttt{mbarrier} \\
  &|& \texttt{CudaCommitGroup} & CUDA \texttt{commit\_group} mechanism \\
  &|& \texttt{CudaClusterSync} & \texttt{barrier.cluster} sync \\
\bottomrule
\end{tabular}
\caption{CUDA Barrier (completion) mechanisms}
\label{fig:BarrierType}
\end{figure}

\begin{figure}[th]
\footnotesize
\centering
\arraycolsep=1.8pt\def\arraystretch{1.0}
\setlength{\tabcolsep}{2pt}
\begin{tabular}{rrll}
\toprule
$\pi_\mathrm{w} : \mathrm{SpecialWindow}$ & $\Coloneqq$ &
  \texttt{Sm90\_tensorMap(swizzle, *smem\_box)} & \texttt{CUtensorMap} parameterized by swizzle mode and box size \\
\bottomrule
\end{tabular}
\caption{CUDA Special Window Types (for Window Statement)}
\label{fig:SpecialWindow}
\end{figure}

\begin{figure}[th]
\footnotesize
\centering
\arraycolsep=1.8pt\def\arraystretch{1.0}
\setlength{\tabcolsep}{2pt}
\begin{tabular}{rrll}
\toprule
$\tau_q : \mathrm{QualTL}$ & $\Coloneqq$ &
  \texttt{cpu\_in\_order\_qual} & Ordinary CPU reads/writes (\textsf{cpu}) \\
  &|& \texttt{cpu\_cuda\_stream\_qual} & Stream-ordered CUDA API calls (\textsf{strm}) \\
  &|& \texttt{cuda\_in\_order\_rmem\_qual} & non-async CUDA instrs' accesses to registers (\textsf{cuda1}) \\
  &|& \texttt{cuda\_in\_order\_ram\_qual} & non-async CUDA instrs' accesses to SMEM/GMEM (\textsf{cuda2}) \\
  &|& \texttt{Sm80\_cp\_async\_qual} & non-bulk \texttt{cp.async} memory accesses (\textsf{Sm80}) \\
  &|& \texttt{tma\_to\_smem\_async\_qual} & \texttt{cp.async.bulk} GMEM reads, SMEM writes (\textsf{tmaS}) \\
  &|& \texttt{tma\_to\_gmem\_async\_qual} & \texttt{cp.async.bulk} SMEM reads, GMEM writes/reduces (\textsf{tmaG}) \\
  &|& \texttt{wgmma\_async\_rmem\_a\_qual} & \texttt{wgmma.mma\_async} reads from $A$ register parameter (\textsf{wgA}) \\
  &|& \texttt{wgmma\_async\_rmem\_d\_qual} & \texttt{wgmma.mma\_async} accesses to $D$ (accumulator) (\textsf{wgD}) \\
  &|& \texttt{wgmma\_async\_smem\_qual} & \texttt{wgmma.mma\_async} access to SMEM, $A$ or $B$ (\textsf{wgS}) \\
  &|& \texttt{wgmma\_zero\_qual} & special case used to model the \texttt{wgmma} \textsf{scale-d} parameter (\textsf{wg0}) \\
\bottomrule
\end{tabular}
\caption{List of qualitative timelines}
\label{fig:QualTL-def}
\end{figure}

\begin{figure}[th]
\footnotesize
\centering
\arraycolsep=1.8pt\def\arraystretch{1.0}
\setlength{\tabcolsep}{2pt}
\begin{tabular}{|r|l|l l|l l|l l l| l l l l|}
\hline
$\tau_s : \mathrm{SyncTL}$ & trnstv? & \textsf{cpu} & \textsf{strm} & \textsf{cuda1} & \textsf{cuda2} & \textsf{Sm80} & \textsf{tmaS} & \textsf{tmaG} & \textsf{wgA} & \textsf{wgD} & \textsf{wgS} & \textsf{wg0} \\
\hline
\texttt{empty\_sync\_tl} &  &  &  &  &  &  &  &  &  &  &  & \\
\texttt{cpu\_in\_order} & y & full &  &  &  &  &  &  &  &  &  & \\
\texttt{cuda\_stream\_sync} & y &  & full & full & full & full & full & full & full & full & full & \\
\texttt{cuda\_in\_order} & y &  & temp. & full & full & temp. & temp. & temp. & temp. & temp. & temp. & temp.\\
\texttt{cuda\_temporal} &  &  & temp. & temp. & temp. & temp. & temp. & temp. & temp. & temp. & temp. & temp.\\
\texttt{Sm80\_cp\_async} &  &  &  &  &  & full &  &  &  &  &  & \\
\texttt{Sm80\_generic} &  &  & temp. & full & full & full & temp. & temp. & temp. & temp. & temp. & temp.\\
\texttt{tma\_to\_smem\_async} &  &  &  &  &  &  & full &  &  &  &  & \\
\texttt{tma\_to\_gmem\_async} &  &  &  &  &  &  &  & full &  &  &  & \\
\texttt{wgmma\_async\_smem} &  &  &  &  &  &  &  &  &  &  & full & \\
\texttt{wgmma\_fence\_1} &  &  &  & full &  &  &  &  & full & full &  & \\
\texttt{wgmma\_fence\_2} &  &  &  &  &  &  &  &  & full & full &  & \\
\texttt{wgmma\_async} &  &  &  &  &  &  &  &  & full & full & full & \\
\texttt{cuda\_async\_proxy} &  &  &  &  &  &  & full & full &  &  & full & \\
\texttt{cuda\_async\_proxy\_wgmma} &  &  &  &  &  &  & full & full & full & full & full & \\
\texttt{cuda\_generic\_and\_async} &  &  & temp. & full & full & full & full & full & temp. & temp. & full & temp.\\
\hline
\end{tabular}
\caption{SyncTL are defined as composition of QualTL.}
\label{fig:SyncTL-def}
\end{figure}

%% file: tex_gpu/9_am_figures.tex
\FloatBarrier
\section{Abstract Machine Conversion and Semantics Definitions}
\label{appendix:am-figures}

A new record Constructor $\mathcal{R}$ is used to define \texttt{RecordRead} and \texttt{RecordMutate} in Figure~\ref{fig:big-step-semantics}.
\textsc{RecordRead} and \textsc{RecordMutate} first evaluate the accessor \(e^{\#}\) and \(e^{\#}_{z}\) to access sets \(W_{e^{\#}}\) and \(W_{e^{\#}_{z}}\) (via \(\Downarrow_{\mathsf{Acc}}\)). They then update the visibility map \(\rho\) at each key \((x,\mathbf{n}) \in W_{e^{\#}}\) using the new visibility record constructor $\mathcal{R}$ defined below: \textsc{RecordRead} adds it to the \emph{read} visibility record set, setting \(\rho'(x,\mathbf{n}).r = \rho(x,\mathbf{n}).r \cup \mathcal{R}(\cdots)\) while leaving \(\rho(x,\mathbf{n}).m\) and \(\rho(x,\mathbf{n}).b\) unchanged; \textsc{RecordMutate} adds it to the \emph{mutate} visibility record set, setting \(\rho'(x,\mathbf{n}).m = \rho(x,\mathbf{n}).m \cup \mathcal{R}(\cdots)\) while preserving \(\rho(x,\mathbf{n}).r\) and \(\rho(x,\mathbf{n}).b\). In both cases, neither the control \(\imath\) nor the store \(\sigma\) changes.

\begin{definition}[New Record Constructor $\mathcal{R}$]
Define $\mathcal{R}$ as follows:
\[
\mathcal{R}:\;
\mathrm{VL}\times\mathrm{Bool}\times \mathrm{QualTL}\times\mathcal{P}(\mathrm{QualTL})\times \Expr^\# \times \rho \times \mathbb{G}
\;\longrightarrow\; \mathcal{P}(\mathbb{R})
\]
\[
\mathcal{R}(\tau_v,t,\tau_q,\tau_q^*,E^\#,\rho, G)\;=\;
\begin{cases}
\{\,(\tau_q,\; S_{G},\; P_{E^\#,\rho})\,\}, & t=\mathsf{true},\\[4pt]
\{\,(\tau_q,\; S_{\{g\}},\; P_{E^\#,\rho}) \mid g\in G\,\}, & t=\mathsf{false},
\end{cases}
\]
where, for any \(H\subseteq\mathbb{G}\), \(S_H\in\mathbb{S}\) and \(P_{E^\#,\rho}\in\mathbb{P}\) are given pointwise by
\[
S_H(g',\tau_q') \;=\;
\max\nolimits_{\mathrm{VL}}\!\left(
\begin{cases}
\mathsf{atomic\_only}, & \tau_q'\in\tau_q^*,\\
\mathsf{invisible}, & \text{otherwise}
\end{cases},
\;
\begin{cases}
\tau_v, & g'\in H \ \wedge\ \tau_q'=\tau_q,\\
\mathsf{invisible}, & \text{otherwise}
\end{cases}
\right),
\]
\[
P_{E^\#,\rho}(z,\mathbf{n})\;=\;
\begin{cases}
\{a\}, & (z,\mathbf{n})\in E^\#\ \text{where}\ (a,\_)=\rho(z,\mathbf{n}).b,  \\
\emptyset, & \text{otherwise.}
\end{cases}
\]
\end{definition}

\texttt{Fence}, \texttt{Arrive}, and \texttt{Await} (Figure~\ref{fig:big-step-semantics}) update $\rho$ solely via $\mathrm{lift}(\rho,\lambda)$ (defined below), which applies $\lambda$ pointwise to both the read and mutate visibility sets.
For \texttt{Fence}$(t,\tau_q^{\mathrm{pre}*},\tau_q^{\mathrm{full}*},\tau_q^{\mathrm{temp}*})$, $\lambda(r)=\mathcal{A}\!\left(r,\,g^{\mathrm{exec}*},\,\tau_q^{\mathrm{full}*},\,\tau_q^{\mathrm{temp}*}\right)$ iff the witness $\mathcal{W}\!\left(t,\,g^{\mathrm{exec}*},\,\tau_q^{\mathrm{pre}*},\,r\right)$ holds, otherwise $\lambda(r)=r$; thus visibility is raised by joining $f\vee f_{\mathrm{aug}}$. 
For \texttt{Arrive}$(t,\tau_q^{\mathrm{pre}*},e_z^{\#*})$, the accessed sets yield a unique home barrier $\{(z,\mathbf{n})\}=\bigcap_i W_{e_{zi}^{\#}}$; guarded by $\mathcal{W}$, $\lambda$ adds the existing arrive count $a$ from $\rho(z,\mathbf{n})$ into $r.p$ at every $(z,\mathbf{n})\in\mathbf{W}$, and home barrier's arrive count is incremented by $(a'+1, w')$.
For \texttt{Await}$(e_z^{\#},\tau_q^{\mathrm{full}*},\tau_q^{\mathrm{temp}*},n)$, with $(a,w)=\rho(z,\mathbf{n}).b$, compute $\mathrm{Max}$ and $\mathrm{New}$ as in the rule; set $\lambda(r)=\mathcal{A}\!\left(r,\,g_{s^\#}(\rho),\,\tau_q^{\mathrm{full}*},\,\tau_q^{\mathrm{temp}*}\right)$ iff some observed arrive count $i\in r.p(z,\mathbf{n})$ satisfies $i\le \mathrm{Max}$, else $\lambda(r)=r$; after $\mathrm{lift}$, update the wait count to $(a',\mathrm{New})$. 
In short, $\mathcal{W}$ decides when augmentation is permitted, $\mathcal{A}$ raises visibility to \texttt{fully\_ordered} or \texttt{temporally\_ordered}.

\begin{definition}[Lift]
For $\lambda:\mathbb{R}\to\mathbb{R}$, define $\mathrm{lift}(\rho,\lambda)$ pointwise by
\[
\mathrm{lift}(\rho,\lambda)(x,\mathbf v) \ =\
\bigl(\{\,\lambda(r)\mid r\in R\,\},\ \{\,\lambda(r)\mid r\in M\,\},\ (a,\ w)\bigr)
\quad\text{where } (R,M,(a,w))=\rho(x,\mathbf v).
\]
\end{definition}

\begin{definition}[Synchronization helpers]
Let $r=(\tau_q^{\mathrm{orig}},\mathcal{s},p)\in\mathbb{R}$ with $\mathcal{s}\in\mathbb{S}$.
For $\mathcal{s}_1,\mathcal{s}_2\in\mathbb{S}$, define the pointwise join
$
(\mathcal{s}_1\vee \mathcal{s}_2)(g,\tau_q)\;=\;\max_{\mathrm{VL}}\{\mathcal{s}_1(g,\tau_q),\,\mathcal{s}_2(g,\tau_q)\}.
$

\emph{(i) Witness $\mathcal{W}$.}
For $t\in\mathrm{Bool}$, $g^{\mathrm{exec}*}\subseteq\mathbb{G}$, and
$\tau_q^{\mathrm{pre}*}\subseteq\mathrm{QualTL}$, set
\[
\mathcal{W}(t, g^{\mathrm{exec}*}, \tau_q^{\mathrm{pre}*}, r)
=
\begin{cases}
\exists g\in g^{\mathrm{exec}*}\,\exists \tau_q\in \tau_q^{\mathrm{pre}*}:\ \mathcal{s}(g,\tau_q)\succeq\texttt{unordered},
& t=\mathsf{true},\\[2pt]
\bigl(\tau_q^{\mathrm{orig}}\in\tau_q^{\mathrm{pre}*}\bigr)\ \wedge\
\exists g\in g^{\mathrm{exec}*}:\ \mathcal{s}(g,\tau_q^{\mathrm{orig}})\succeq\texttt{unordered},
& t=\mathsf{false}.
\end{cases}
\]

\emph{(ii) Augment $\mathcal{A}$.}
Given $g^{\mathrm{exec}*}\subseteq\mathbb{G}$ and
$\tau_q^{\mathrm{full}*},\tau_q^{\mathrm{temp}*}\subseteq\mathrm{QualTL}$, define 
\[
\mathcal{A}(r, g^{\mathrm{exec}*}, \tau_q^{\mathrm{full}*}, \tau_q^{\mathrm{temp}*})
\;=\; \bigl(\tau_q^{\mathrm{orig}},\, \mathcal{s}\vee \mathcal{s}_{\mathrm{aug}},\, p\bigr),
\]
\[
\mathcal{s}_{\mathrm{aug}}(g,\tau_q)=
\begin{cases}
\texttt{fully\_ordered}, & g\in g^{\mathrm{exec}*}\ \wedge\ \tau_q\in\tau_q^{\mathrm{full}*},\\
\texttt{temporally\_ordered}, & g\in g^{\mathrm{exec}*}\ \wedge\ \tau_q\in\tau_q^{\mathrm{temp}*}\setminus\tau_q^{\mathrm{full}*},\\
\texttt{invisible}, & \text{otherwise.}
\end{cases}
\]
\end{definition}

\begin{figure}[t]
\centering
\footnotesize
\setlength{\tabcolsep}{4pt}
\begin{tabular}{ m{5.2cm} m{4.6cm} }
\toprule
\emph{Data expressions $e \in \mathrm{DExpr}$} & $\mathcal{P}(\Expr^\#)$ \\ \midrule
$d$ & $\emptyset$ \\[2pt]
$x$ & $\{\,x[\langle \rangle]\,\}$ \\[2pt]
$x[w^*]$ & $\{\,x[w^*] \,\}$ \\[2pt]
$e_1 \; \mathrm{op} \; e_2$ \;\; ($\mathrm{op} \in \{+, -, *, /\}$) & $T_e(e_1)\; \cup\; T_e(e_2)$ \\ \midrule
\multicolumn{2}{l}{\emph{Barrier expressions $e_{\mathrm{z}} \in \mathrm{ZExpr}$}} \\ \midrule
$z$ & $\{\,z[\langle \rangle]\,\}$ \\[2pt]
$z[w^*]$ & $\{\,z[w^*] \,\}$ \\
\bottomrule
\end{tabular}
\caption{
Expression conversion
$T_e : \mathrm{DExpr} \cup \mathrm{ZExpr}  \to \mathcal{P}(\Expr^\#)$ collects the distinct reads in an expression and returns them as a set of AMIR reads ($\Expr^\#$).
}
\label{fig:expression_conversion}
\end{figure}

\begin{figure}[t]
\centering
\footnotesize
\setlength{\tabcolsep}{6pt}
\begin{tabular}{@{}p{3.6cm} p{10cm}@{}}
\toprule
\textbf{Type} & \textbf{Definition} \\
\midrule

$T_\mathrm{smem} : \Expr \to \mathcal{P}(\Expr^\#)$
&
$T_\mathrm{smem}(e) \;\triangleq\; \{\, e^\# \in T_e(e) \mid \mathrm{isSmem}(e^\#) \,\}.$
\\[4pt]

$T_\mathrm{ex} : \Expr \to \mathcal{P}(\Expr^\#)$
&
$T_\mathrm{ex}(e) \;\triangleq\; \{\, e^\# \in T_e(e) \mid \neg\,\mathrm{isEx}(e^\#) \,\}.$
\\[6pt]

$\mathrm{isEx} : \Expr^\# \to \mathsf{Bool}$
&
$\mathrm{isEx}(e^\#) \;\triangleq\; \texttt{true} \text{ iff $e^\#$ is annotated as sync‑exempt memory; \texttt{false} otherwise.}$
\\[6pt]

$\mathrm{isSmem} : \Expr^\# \to \mathsf{Bool}$
&
$\mathrm{isSmem}(e^\#) \;\triangleq\; \texttt{true} \text{ iff $e^\#$ is annotated as shared memory; \texttt{false} otherwise.}$
\\[6pt]

$T_\mathrm{tr} : \SyncTL \to \mathsf{Bool}$
&
$T_\mathrm{tr}(\tau_s) \;\triangleq\; \texttt{true} \text{ iff } \tau_s \text{ is transitive; \texttt{false} otherwise.}$
\\[6pt]

$T_\mathrm{qfull} : \SyncTL \to \mathcal{P}(\QualTL)$
&
$T_\mathrm{qfull}(\tau_s) \;\triangleq\; \{\, \tau_q' \in \QualTL \mid \tau_q' \text{ annotated as ``full'' in }\tau_s \,\}.$
\\[6pt]

$T_\mathrm{qtmp} : \SyncTL \to \mathcal{P}(\QualTL)$
&
$T_\mathrm{qtmp}(\tau_s) \;\triangleq\; \{\, \tau_q' \in \QualTL \mid \tau_q' \text{ annotated as ``temporal'' or ``full'' in }\tau_s \,\}.$
\\[6pt]


$T_\mathrm{qarr} : \SyncTL \to \QualTL$
&
$T_\mathrm{qarr}(\tau_s) \;\triangleq\; 
\begin{cases}
\texttt{Sm80\_cp\_async\_qual}, & \text{if } \texttt{Sm80\_cp\_async\_qual} \in T_\mathrm{qfull}(\tau_s),\\
\texttt{cuda\_in\_order\_qual}, & \text{otherwise.}
\end{cases}$
\\[10pt]

$T_{\mathrm{qin}} : \Expr^\# \to \QualTL$
&
$T_{\mathrm{qin}}(e^\#) \;\triangleq\;
\begin{cases}
\texttt{cpu\_in\_order\_qual},     & \text{if $e^\#$ is host/CPU scoped}, \\
\texttt{cuda\_in\_order\_rmem\_qual}, & \text{if } e^\# \text{ is register memory},\\
\texttt{cuda\_in\_order\_ram\_qual},  & \text{otherwise.}
\end{cases}$
\\

\bottomrule
\end{tabular}

\caption{Helper functions used by Fig.~\ref{fig:statement_conversion} statement conversion.
All sets are finite.
}
\label{fig:statement_conversion_helpers}
\end{figure}

\begin{figure}[t]
\centering
\footnotesize
\setlength{\tabcolsep}{4pt}
\begin{tabular}{ m{3.6cm} m{9.5cm} }
\toprule
\textbf{Case for $g.p_i$} & \textbf{Abstract Machine IR} \\
\midrule
Non-atomic \emph{read-only} &
$\displaystyle
\bigseq_{e^\# \in T_\mathrm{ex}(e_i)} \Bigl(
\texttt{CheckMutates}(\texttt{full\_ordered}, T_\mathrm{cvg}(g.p_i), e^\#,~ T_\mathrm{ext}(g.p_i)) \texttt{;}$\\
$\hspace{5.1cm}\texttt{RecordRead}(\vpost, T_\mathrm{cvg}(g.p_i), e^\#,~ T_\mathrm{init}(g.p_i),~ \emptyset, T_e(e_\mathrm{z}))
\Bigr)
$ \\
\midrule
Non-atomic \emph{non-read-only} &
$\displaystyle
\bigseq_{e^\# \in T_\mathrm{ex}(e_i)} \Bigl(
\texttt{CheckMutates}(\tau_v^\mathrm{pre}, T_\mathrm{cvg}(g.p_i), e^\#,~ T_\mathrm{ext}(g.p_i)) \texttt{;}$\\
$\hspace{5.1cm}\texttt{CheckReads}(\VLtmp, T_\mathrm{cvg}(g.p_i), e^\#,~ T_\mathrm{ext}(g.p_i)) \texttt{;}$\\
$\hspace{5.1cm}\texttt{ClearReads}(e^\#) \texttt{;}$\\
$\hspace{5.1cm}\texttt{ClearMutates}(e^\#) \texttt{;}$\\
$\hspace{5.1cm}\texttt{RecordMutate}(\tau_v^\mathrm{post}, T_\mathrm{cvg}(g.p_i), e^\#,~ T_\mathrm{init}(g.p_i),~ \emptyset, T_e(e_\mathrm{z}))
\Bigr)
$ \\
& \hspace*{1em}where $\tau_v^\mathrm{pre} = \VLtmp$ if $g.p_i$ is write-only, else $\VLfull$.\\
\midrule
\emph{Atomic} &
$\displaystyle
\bigseq_{e^\# \in T_\mathrm{ex}(e_i)} \Bigl(
\texttt{CheckMutates}(\VLatom, T_\mathrm{cvg}(g.p_i), e^\#,~ T_\mathrm{ext}(g.p_i)) \texttt{;}$\\
$\hspace{5.1cm}\texttt{CheckReads}(\VLtmp, T_\mathrm{cvg}(g.p_i), e^\#,~ T_\mathrm{ext}(g.p_i)) \texttt{;}$\\
$\hspace{5.1cm}\texttt{ClearReads}(e^\#) \texttt{;} \textit{(note lack of ClearMutates)}$\\
$\hspace{5.1cm}\texttt{RecordMutate}(\tau_v^\mathrm{post}, T_\mathrm{cvg}(g.p_i), e^\#,~ T_\mathrm{init}(g.p_i), T_\mathrm{atom}(g.p_i), T_e(e_\mathrm{z}))
\Bigr)
$ \\
\bottomrule
\end{tabular}
\caption{
Argument conversion
$T_\mathrm{arg}(g.p_i, e_i, e_z)$. $\vpost = \texttt{unordered}$ if $T_\mathrm{ooo}(g.p_i)$, else $\texttt{full\_ordered}$.
}
\label{fig:targ_conversion}
\end{figure}

\begin{figure}[t]
\centering
\footnotesize
\setlength{\tabcolsep}{6pt}
\begin{tabular}{@{}p{3.6cm} p{9cm}@{}}
\toprule
\textbf{Type} & \textbf{Definition} \\
\midrule

$T_\mathrm{ext} : \Expr \to \mathcal{P}(\QualTL)$
&
$T_\mathrm{ext}(e) \;\triangleq\; \{\, \tau_q \in \QualTL \mid \tau_q \in e.\texttt{ext\_qual\_tl} \,\}.$
\\[6pt]

$T_\mathrm{atom} : \Expr \to \mathcal{P}(\QualTL)$
&
$T_\mathrm{atom}(e) \;\triangleq\; \{\, \tau_q \in \QualTL \mid \tau_q \in e.\texttt{atomic\_qual\_tl} \,\}.$
\\[6pt]

$T_\mathrm{init} : \Expr \to \QualTL$
&
$T_\mathrm{init}(e) \;\triangleq\; e.\texttt{initial\_qual\_tl}.$
\\[6pt]

$T_\mathrm{ooo} : \Expr \to \mathsf{Bool}$
&
$T_\mathrm{ooo}(e) \;\triangleq\; e.\texttt{out\_of\_order}$
\\[6pt]

$T_\mathrm{cvg} : \Expr \to \mathsf{Bool}$
&
$T_\mathrm{cvg}(e) \;\triangleq\; e.\texttt{convergent}$
\\

\bottomrule
\end{tabular}

\caption{Helper functions used by Fig.~\ref{fig:targ_conversion} argument conversion.
}
\label{fig:targ_conversion_helpers}
\end{figure}

\begin{figure}[t]
\centering
\small
\begin{minipage}[h]{\linewidth}
\centering
\begin{mathpar}

\inferrule*[right=LoopStep]{
  \langle c_{\mathrm{lo}}, \sigma \rangle \Downarrow_{\mathbb{Z}} m \quad
  \langle c_{\mathrm{hi}}, \sigma \rangle \Downarrow_{\mathbb{Z}} n \quad
  m < n \quad
  \langle s, \imath, \sigma[y \mapsto m], \rho \rangle \Downarrow \imath', \sigma',\rho' \\
  \langle \text{for } y \text{ in } (m{+}1) \dots n \text{ do } s^\#, \imath', \sigma', \rho' \rangle \Downarrow \imath'', \sigma'', \rho''
}{
  \langle \text{for } y \text{ in } \texttt{loop}(c_{\mathrm{lo}}, c_{\mathrm{hi}}) \text{ do } s^\#, \imath, \sigma, \rho \rangle
  \Downarrow \imath'', \sigma'', \rho''
}

\inferrule*[right=LoopDone]{
  \langle c_{\mathrm{lo}}, \sigma \rangle \Downarrow_{\mathbb{Z}} m \quad
  \langle c_{\mathrm{hi}}, \sigma \rangle \Downarrow_{\mathbb{Z}} n \quad
  m \geq n
}{
  \langle \text{for } y \text{ in } \texttt{loop}(c_{\mathrm{lo}}, c_{\mathrm{hi}}) \text{ do } s^\#, \imath, \sigma, \rho \rangle
  \Downarrow \imath, \sigma, \rho
}

\inferrule*[right=Seq]{
 \langle s_{1}^\#,\imath,\sigma,\rho\rangle\Downarrow\imath',\sigma',\rho'\,\and\,
 \langle s_{2}^\#,\imath',\sigma',\rho'\rangle\Downarrow\imath'',\sigma'',\rho''
}{
    \langle s_{1}^\#;\,s_{2}^\#,\imath,\sigma,\rho\rangle\Downarrow\imath'',\sigma'',\rho''
}

\inferrule*[right=InctaskID]{
  \imath' = \imath + 1
}{
  \langle \text{InctaskID}, \imath, \sigma, \rho \rangle
  \Downarrow \imath', \sigma, \rho
}

\inferrule*[right=IF-True]{
  \langle b, \sigma \rangle \Downarrow_{\mathsf{Bool}} \text{true} \quad
  \langle s^\#, \imath, \sigma, \rho \rangle \Downarrow \imath', \sigma', \rho'
}{
  \langle \text{if } b \text{ then } s^\#, \imath, \sigma, \rho \rangle
  \Downarrow \imath', \sigma', \rho'
}

\inferrule*[right=IF-False]{
  \langle b, \sigma \rangle \Downarrow_{\mathsf{Bool}} \text{false}
}{
  \langle \text{if } b \text{ then } s^\#, \imath, \sigma, \rho \rangle
  \Downarrow \imath, \sigma, \rho
}

\end{mathpar}
\end{minipage}

\caption{
Big-step semantics for a loop, sequencing, an explicit task–ID increment, and a guard.
}
\label{fig:big-step-semantics-loops}
\end{figure}

\begin{figure}[t]
\centering
\small
\begin{minipage}[h]{\linewidth}
\centering
\begin{mathpar}

\inferrule*[right=Alloc]{
  c^*=(c_1,\dots,c_k) \quad
  \big\langle c_i, \sigma \big\rangle \Downarrow_{\mathbb{Z}} n_i\ \text{ for } i=1..k \\
  I \;=\; \{\, (v_1,\dots,v_k) \mid 0 \le v_i < n_i \,\} \\
  \rho' \;=\; \rho\bigl[(x,\mathbf{n}) \mapsto (\emptyset,\emptyset,(0,0))\bigr]_{\mathbf{n}\in I}
}{
  \langle \text{Alloc(}x[c^*]\text{)},\, \imath,\, \sigma,\, \rho \rangle
  \Downarrow \imath, \sigma, \rho'
}

\inferrule*[right=ClearReads]{
\langle e^\#, \sigma \rangle \Downarrow_{\mathsf{Acc}} W_{e^\#} 
\\
\rho' \;=\; \rho\bigl[(x,\mathbf{n}) \mapsto (\,\emptyset,\, \rho(x, \mathbf{n}).m,\, \rho(x, \mathbf{n}).b\,)\bigr]_{\ (x, \mathbf{n}) \in W_{e^\#}}
}{
  \bigl\langle \text{ClearReads(}e^\#\text{)},\ \imath,\ \sigma,\ \rho \bigr\rangle
  \Downarrow\ \imath,\ \sigma,\ \rho'
}

\inferrule*[right=ClearMutates]{
\langle e^\#, \sigma \rangle \Downarrow_{\mathsf{Acc}} W_{e^\#} 
\\
\rho' \;=\; \rho\bigl[(x,\mathbf{n}) \mapsto (\, \rho(x, \mathbf{n}).r, \emptyset, \rho(x, \mathbf{n}).b\,)\bigr]_{\ (x, \mathbf{n}) \in W_{e^\#}}
}{
  \bigl\langle \text{ClearMutates(}e^\#\text{)},\ \imath,\ \sigma,\ \rho \bigr\rangle
  \Downarrow\ \imath,\ \sigma,\ \rho'
}

\inferrule*[right=RecordRead]{
  \langle e^\#, \sigma \rangle \Downarrow_{\mathsf{Acc}} W_{e^\#} 
  \quad \langle e^{\#}_z, \sigma \rangle \Downarrow_{\mathsf{Acc}} W_{e^{\#}_z} 
  \\
  \quad   \rho' \;=\; \rho\bigl[(x,\mathbf{n}) \mapsto (\,\rho(x, \mathbf{n}).r \cup \mathcal{R}(\tau_v,\, t,\, \tau_q,\, \tau_q^*,\, W_{e^{\#}_z}, \rho, g_{s^\#}(\sigma, \imath)),\, \rho(x, \mathbf{n}).m,\, \rho(x, \mathbf{n}).b\,)\bigr]_{\ (x, \mathbf{n}) \in W_{e^\#}}
}{
  \langle \text{RecordRead(}\tau_v,\, t,\, e^\#,\, \tau_q,\, \tau_q^*,\, e_z^{\#}\text{)},\, \imath,\, \sigma,\, \rho \rangle
  \Downarrow \imath, \sigma, \rho'
}

\inferrule*[right=RecordMutate]{
  \langle e^\#, \sigma \rangle \Downarrow_{\mathsf{Acc}} W_{e^\#} 
  \quad \langle e^{\#}_z, \sigma \rangle \Downarrow_{\mathsf{Acc}} W_{e^{\#}_z} 
  \\
  \quad   \rho' \;=\; \rho\bigl[(x,\mathbf{n}) \mapsto (\,\rho(x, \mathbf{n}).r,\, \rho(x, \mathbf{n}).m \cup \mathcal{R}(\tau_v,\, t,\, \tau_q,\, \tau_q^*,\, W_{e^{\#}_z}, \rho, g_{s^\#}(\sigma, \imath)),\, \rho(x, \mathbf{n}).b\,)\bigr]_{\ (x, \mathbf{n}) \in W_{e^\#}}
}{
  \langle \text{RecordMutate(}\tau_v,\, t,\, e^\#,\, \tau_q,\, \tau_q^*,\, e_z^{\#}\text{)},\, \imath,\, \sigma,\, \rho \rangle
  \Downarrow \imath, \sigma, \rho'
}

\newcommand{\ExecG}{g_{s^\#}(\sigma,\imath)}

\inferrule*[right=Fence]{
  \lambda\ =\ \bigl(r \mapsto \mathcal{A} (r,\ExecG,\tau_q^{\mathrm{full}*},\tau_q^{\mathrm{temp}*}) \text{ if } \mathcal{W}(t,\ExecG,\tau_q^{\mathrm{pre}*},r ) \text{ else } r\bigr)
  \\
}{
  \bigl\langle \texttt{Fence}(t,\tau_q^{\mathrm{pre}*},\tau_q^{\mathrm{full}*},\tau_q^{\mathrm{temp}*}),\ \imath,\ \sigma,\ \rho \bigr\rangle
  \Downarrow\ \imath,\ \sigma,\ \mathrm{lift}(\rho, \lambda)
}

\inferrule*[right=Arrive]{
  \langle \langle e_{z1}^{\#}, \cdots, e_{zk}^\# \rangle, \sigma \rangle \Downarrow_{\mathsf{Acc}} W_{e_{z1}^{\#}}, \cdots, W_{e_{zk}^{\#}}
  \quad \{ (z, \mathbf{n}) \} = \bigcap_{i=1}^k W_{e_{zi}^{\#}}
  \quad \mathbf{W} = \bigcup_{i=1}^k W_{e_{zi}^{\#}}
  \\
  \lambda\ =\ \bigl(r \mapsto \bigl(r.q,\, r.\mathcal{s},\, r.p[(z, \mathbf{n}) \mapsto r.p(z, \mathbf{n}) \cup \{a\}]_{z, \mathbf{n} \in \mathbf{W}}^{(a, \_) = \rho(z, \mathbf{n})} \bigr) \, \text{ if } \, \mathcal{W}(t,\ExecG,\tau_q^{\mathrm{pre}*},r ) \, \text{ else }\, r \, \bigr)
  \\
  \rho' = \mathrm{lift}(\rho, \lambda)
  \\
  (R',M',(a',w'))\;=\;\rho'(z,\mathbf{n})\qquad
  \rho''\;=\;\rho'[(z,\mathbf{n})\mapsto (R',M',(a'+1,w'))]
}{
  \bigl\langle \texttt{Arrive}(t,\tau_q^{\mathrm{pre}*}, e_z^{\#*}),\ \imath,\ \sigma,\ \rho \bigr\rangle
  \Downarrow\ \imath,\ \sigma,\ \rho''
}

\inferrule*[right=Await]{
  \langle e^{\#}_z, \sigma \rangle \Downarrow_{\mathsf{Acc}} \{(z, \mathbf{n})\}
  \quad
  (a,w) = \rho(z,\mathbf{n}).b
  \\
  \mathrm{Max} \;\triangleq\;
  \text{if } n \ge 0 \text{ then } a-(n+1) \text{ else } w+n+1
  \\
  \mathrm{New} \;\triangleq\;
  \text{if } n \ge 0 \text{ then } \max\!\bigl(w,\ \mathrm{Max} +1\bigr) \text{ else } w+1
  \\
  \lambda \;=\; \Bigl(r \mapsto
    \mathcal{A}\bigl(r,\ g_{s^\#}(\rho),\ \tau_q^{\mathrm{full}*},\ \tau_q^{\mathrm{temp}*}\bigr)
    \ \text{if }\ \exists i.\ i \le \mathrm{Max} \wedge i \in r.p(z,\mathbf{n})
    \ \text{ else }\ r\Bigr)
  \\
  \rho' = \mathrm{lift}(\rho, \lambda)
  \qquad
  (R', M', (a', \_))\;=\;\rho'(z,\mathbf{n})
  \\
  \rho''\;=\;\rho'[(z,\mathbf{n})\mapsto (R', M', (a', \mathrm{New}))]
}{
  \bigl\langle \texttt{Await}(e_z^{\#},\ \tau_q^{\mathrm{full}*},\ \tau_q^{\mathrm{temp}*},\ n),\ \imath,\ \sigma,\ \rho \bigr\rangle
  \Downarrow\ \imath,\ \sigma,\ \rho''
}

\end{mathpar}
\end{minipage}

\caption{The big-step operational semantics for Abstract Machine statements with sync environment updates.}
\label{fig:big-step-semantics}
\end{figure}